\documentclass[fleqn,10pt]{olplainarticle}
\usepackage{float}
\usepackage{graphicx}
\usepackage{epstopdf, epsfig}
\usepackage{todonotes}
\usepackage{subcaption}
\usepackage{amsmath}
\usepackage{amssymb}
\usepackage{color}
\usepackage{placeins}
\usepackage{natbib}

\title{Closing the loop: nonlinear Taylor vortex flow through the lens of resolvent analysis}

\author[1]{Benedikt Barthel}
\author[2]{Xiaojue Zhu}
\author[3]{Beverley J. McKeon}
\affil[1]{Graduate Aerospace Laboratories, California Institute of Technology,
Pasadena, CA 91125, USA}
\affil[2]{Center of Mathematical Sciences and Applications, and School of Engineering and Applied Sciences, Harvard University, Cambridge, MA 02138, USA}
\affil[1]{Graduate Aerospace Laboratories, California Institute of Technology,
Pasadena, CA 91125, USA}

\keywords{Taylor-Couette flow, low-dimensional models, nonlinear instability, transition to turbulence}

\begin{abstract}
We present an optimization-based method to efficiently calculate accurate nonlinear models of Taylor vortex flow. We use the resolvent formulation of \citet{mckeon_critical-layer_2010} to model these Taylor vortex solutions by treating the nonlinearity not as an inherent part of the governing equations but rather as a triadic constraint which must be satisfied by the model solution. We exploit the low rank linear dynamics of the system to calculate an efficient basis for our solution, the coefficients of which are then calculated through an optimization problem where the cost function to be minimized is the triadic consistency of the solution with itself as well as with the input mean flow. Our approach constitutes, what is to the best of our knowledge, the first fully nonlinear and self-sustaining, resolvent-based model described in the literature. We compare our results to direct numerical simulation of Taylor Couette flow at up to five times the critical Reynolds number, and show that our model accurately captures the structure of the flow. Additionally, we find that as the Reynolds number increases the flow undergoes a fundamental transition from a classical weakly nonlinear regime, where the forcing cascade is strictly down scale, to a fully nonlinear regime characterized by the emergence of an inverse (up scale) forcing cascade. Triadic contributions from the inverse and traditional cascade destructively interfere implying that the accurate modeling of a certain Fourier mode requires knowledge of its immediate harmonic and sub-harmonic. We show analytically that this finding is a direct consequence of the structure of the quadratic nonlinearity of the governing equations formulated in Fourier space. Finally, we show that using our model solution as an initial condition to a higher Reynolds number DNS significantly reduces the time to convergence.
\end{abstract}

\begin{document}
\maketitle

\section{Introduction}
Taylor-Couette flow (TCF), the flow between two concentric and independently rotating cylinders, is one of the canonical problems in fluid mechanics, and a paradigm for the study of linear stability, pattern formation, and rotationally driven turbulence. From the original investigations of \citet{taylor_viii_1923} to the pioneering experiments of \citet{coles_transition_1965}, recent theoretical analyses \citep{gebhardt_taylor-couette_1993,jones_nonlinear_1981,maretzke_transient_2014}, high Reynolds number simulations \citep{ostilla_optimal_2013,ostilla-monico_exploring_2014,grossmann_highreynolds_2016,sacco_dynamics_2019}, and experimental investigation \citep{van_gils_twente_2011,huisman_multiple_2014,van_gils_optimal_2012}, TCF has remained a problem of interest for most of the last century. Perhaps the most well known characteristic of TCF is the incredibly rich array of stable flow states that exist over a range of geometries and relative rotation rates of the inner and outer cylinders \citep{coles_transition_1965,andereck_flow_1986}. We consider the case of pure inner cylinder rotation, for which the problem is parameterized by the ratio of the inner and outer radii, $\eta \equiv r_i/r_o$, and a single Reynolds number: $R$.  In this case the laminar velocity profile becomes linearly unstable at a critical Reynolds number: $R_c \sim \mathcal{O} (10^2)$. A centrifugal instability leads to the formation of a periodic array of toroidal vortex structures known as Taylor vortices or Taylor rolls. For a given geometry, these steady, axisymmetric Taylor vortices are stable and exist for some range of Reynolds number in what is known as the Taylor Vortex flow (TVF) regime. As the rotation rate of the inner cylinder is increased further, the Taylor vortices experience a secondary instability giving rise to azimuthally traveling waves whose phase speed is determined by the geometry but whose azimuthal periodicity is not unique \citep{coles_transition_1965}. This regime is known as Wavy Vortex flow (WVF) and is characterized by being time periodic in a stationary reference frame but steady in a frame corotating with the traveling wave \citep{marcus_simulation_1984}. WVF is again stable for some range of Reynolds numbers before the traveling waves themselves become unstable, and a second temporal frequency arises causing the traveling waves to become modulated in space and time in what is known as modulated wavy vortex flow (MWVF). As the driving of the inner cylinder is increased further still, the flow becomes disordered and begins to transition to turbulence. However, while the main sequence of transitions from laminar flow to the bifrucation to TVF at $R_c \approx 100$, to WVF, to MWVF, and finally on to turbulence at $R \approx 1000$ occurs over a relatively narrow range of Reynolds numbers, the large scale Taylor vortices are present up to $R \sim \mathcal{O} (10^5)$ \citep{grossmann_highreynolds_2016}. 

In recent years there has been renewed interest \citep{dessup_self-sustaining_2018,sacco_dynamics_2019} in the dynamics of TVF and WVF as a model system to study the self-sustaining process (SSP) proposed by \citet{waleffe_self-sustaining_1997}. The SSP consists of streamwise rolls which advect the mean shear giving rise to streaks of streamwise velocity, which become unstable to wave-like disturbances, which in turn nonlinearly interact to sustain the rolls \citep{waleffe_self-sustaining_1997,hamilton_regeneration_1995}. \citet{dessup_self-sustaining_2018} performed direct numerical simulations (DNS) to show that the mechanism of transition from TVF to WVF follows the same path as described in the SSP: the traveling waves of WVF arise due to an instability of the streamwise velocity component (streaks) of TVF with the cross-stream velocity (rolls) playing a negligible role in the instability mechanism. \citet{sacco_dynamics_2019} extended this line of study to higher Reynolds numbers and showed that despite their origin as a centrifugal instability, turbulent Taylor vortices are preserved in the limit of vanishing curvature and are thus not dependent on a rotational effects, and rather are sustained through a nonlinear feedback loop between the rolls and streaks. 

The SSP is believed to be one of the building blocks of turbulence and thus the study of self sustaining solutions of the Navier-Stokes equations (NSE), known as \textit{exact coherent states} (ECS) has been of great interest to researchers since they were first discovered by \citet{nagata_three-dimensional_1990}. The field has grown immensely since in the intervening years and we make no attempt to summarize it all here in the interests of brevity. Of primary interest is the observation that many of these solutions resemble streamwise elongated vortices and streaks, and thus resemble structures observed in experiments and simulations of turbulent flows \citep{beaume_reduced_2015}. This has led to the idea that ECS make up the phase space skeleton of turbulence, and that the observation of these structures indicates the turbulent trajectory passing by one of these ECS solutions. 

Due to the significant mathematical simplification, it is useful to model these elongated structures as being infinitely long in their streamwise extent. \citet{illingworth_streamwise-constant_2020} studied the linear amplification mechanism of such streamwise invariant structures to identify the most amplified spanwise length scales in both channel and plane Couette flow (PCF), and found that the latter was far more efficient in amplifying these structures. However no structures actually observed in channel or plane Couette flow are truly invariant in the streamwise direction. Additionally, ECS are generally unstable; while the turbulent trajectory may visit these states, they do not actually persist in nature. TVF is thus a valuable test case to study the nonlinear dynamics sustaining ECS in general since it is in fact a stable solution observed in experiment and due to the cylindrical geometry is exactly streamwise constant.

 We aim to help bridge the gap between linear stability theory, which accurately predicts the genesis of Taylor vortices, and higher Reynolds numbers where Taylor vortices are sustained by fully nonlinear mechanisms as shown by \citet{sacco_dynamics_2019}. The resolvent formulation of the NSE, which interprets the nonlinear term in the NSE as a forcing to the linear dynamics, offers a natural path from linear to nonlinear analysis and as such will form the basis of our modeling efforts. Resolvent analysis has historically been successful in exploiting the linear dynamics to identify structures in turbulence, which is by nature a nonlinear phenomenon \citep{mckeon_critical-layer_2010}. However, recently attempts have been made to explicitly characterize and quantify the influence of nonlinear dynamics within the resolvent framework. For example, \citet{moarref_low-order_2014} and \citet{mcmullen_interaction_2020} used convex optimization to compute reduced order representations of turbulent statistics and \citet{rigas_nonlinear_2021} studied solutions of the Harmonic-Balanced Navier-Stokes equations to identify optimal nonlinear mechanisms leading to boundary layer transition. Additionally \citet{morra_colour_2021} and \citet{nogueira_forcing_2021} have directly computed the nonlinear forcing statistics for minimal channel and Couette flow respectively and analyzed the efficacy of low rank resolvent reconstructions in capturing the relevant dynamics.

 In this work we use optimization-based methods not only to capture statistics, but to explicitly model the self-sustaining nonlinear system. To the best of these authors' knowledge this work is the first example in the literature of ``\textit{closing the resolvent loop}''  where the feedback loop formulation of the NSE introduced by \citet{mckeon_critical-layer_2010} is used to generate solutions to the fully nonlinear NSE.  In \S\ref{sec:math} we outline the resolvent formulation of the NSE and our optimization-based model.  In \S\ref{sec:results} we present the results of our model, compare them to DNS, and present a detailed analysis of the nonlinear dynamics. We discuss applications of our results to DNS of turbulent TCF in \S\ref{sec:DNSic} and conclude with some final remarks in \S\ref{sec:conclusion}.

\section{Mathematical Description}\label{sec:math}
We study the flow of an incompressible Newtonian fluid with kinematic viscosity $\nu$ between two concentric cylinders using the NSE in cylindrical coordinates,
\begin{equation} \label{NSE}
    \frac{\partial \tilde{\boldsymbol{u}}}{\partial t} + \tilde{\boldsymbol{u}} \cdot\nabla \tilde{\boldsymbol{u}} - \frac{1}{R}\nabla^2 \tilde{\boldsymbol{u}} -\nabla \tilde{p} = 0
\end{equation}
\begin{equation}\label{continuity}
    \nabla \cdot \tilde{\boldsymbol{u}}=0
\end{equation}
on the domain $r\in [r_i,r_o]$, $\theta \in [0, 2\pi]$, $z\in(-L_z/2,L_z/2)$. We will consider the case where the outer cylinder is held fixed while the inner cylinder rotates with a prescribed azimuthal speed $U_i$. The equations are nondimensionalized using the gap width $d\equiv r_o-r_i$ and the azimuthal velocity of the inner cylinder $U_i$. The  Reynolds number is defined as $R\equiv U_i d/\nu$. In these nondimensional variables the limits of the radial domain are given as a function of the radius ratio $\eta$ by $r_i=\eta/(1-\eta)$ and $r_o = 1/(1-\eta)$. Throughout this work we fix $\eta = 0.714$. This $\eta$ was chosen to allow for comparison to past studies such as \citet{ostilla_optimal_2013} and since it allows for a larger range of Reynolds numbers for which TVF is a stable solution of (\ref{NSE}). 

We decompose the state $[\tilde{\mathbf{u}},\tilde{p}]=[\tilde{u}_\theta,\tilde{u}_r,\tilde{u}_z,p]$ into a mean and fluctuating component,
\begin{equation}
    [\tilde{\mathbf{u}}(r,\theta,z,t),\tilde{p}(r,\theta,z,t)] = [ \mathbf{\overline{U}},\overline{P}(r)] + [\mathbf{u}(r,\theta,z,t),p(r,\theta,z,t)]
\end{equation}
with
\begin{equation}
    \overline{(\cdot)}\equiv \lim_{T,L\to\infty} \frac{1}{2\pi T L} \int_{0}^{2\pi}\int_{0}^{L}\int_{0}^{T} (\cdot)\,d\theta\, dz\, dt,
\end{equation}
which upon substitution into(\ref{NSE}) and averaging over $\theta, z$, and $t$ results in the mean momentum equation 
\begin{equation} \label{RANS}
     \mathbf{\overline{U}} \cdot\nabla \mathbf{\overline{U}} - \frac{1}{R}\nabla^2 \mathbf{\overline{U}} -\nabla \overline{P} = -\overline{\mathbf{u} \cdot\nabla \mathbf{u}}
\end{equation}
\begin{equation}
    \nabla \cdot \mathbf{\overline{U}} = 0.
\end{equation}
 Subtracting (\ref{RANS}) from the full NSE then results in a governing equation for the fluctuations, where we have grouped those terms which are nonlinear in the fluctuations on the right hand sides in anticipation of the following analysis. 
\begin{equation} \label{pNSE}
    \frac{\partial \mathbf{u}}{\partial t} + \mathbf{\overline{U}} \cdot\nabla \mathbf{u} + \mathbf{u} \cdot\nabla \mathbf{\overline{U}}- \frac{1}{R}\nabla^2 \mathbf{u} -\nabla p = -\left(\mathbf{u} \cdot\nabla \mathbf{u}-\overline{\mathbf{u} \cdot\nabla \mathbf{u}}\right)
\end{equation}
\begin{equation}\label{pcont}
    \nabla \cdot \mathbf{u} = 0
\end{equation}
\begin{equation}
    \mathbf{\overline{U}}|_{r_i} =  \hat{e}_{\theta}, ~\ \mathbf{\overline{U}}|_{r_o} =  \mathbf{u}|_{r_i}=\mathbf{u}|_{r_o} = \mathbf{0}
\end{equation}

\subsection{Direct Numerical Simulation}
In order to validate our model solution we perform DNS of TCF  for a range of Reynolds number, $100<R<2000$, for a radius ratio $\eta = 0.714$ and an aspect ratio $L_z/d = 12$. However, our analysis is focused primarily on the cases $R = 100,200,$ and $400$. The details of the numerical method can be found in \cite{verzicco_finite-difference_1996,van_der_poel_pencil_2015,zhu_afid-gpu_2018}, and the details of the simulations performed in this work are summarized in table \ref{tab:DNS}.
\begin{table}
    \centering
    \begin{tabular}{c|c|c|c|c}
     $R$    &  $N_r$  & $N_{\theta}$ & $N_z$ & DOF $(N_r\times N_{\theta}\times N_z)$ \\ 
     \hline
        100 & 101 & 768 & 512 & 39,714,816  \\
        200 &  101 & 768 & 512 & 39,714,816\\
        400 &  129 & 768 & 512  & 50,724,864\\
        650 &  129 & 768 & 512  & 50,724,864\\
        1000 &  193 & 1024 & 640  & 126,484,480\\
        2000 &  193 & 1024 & 640  & 126,484,480\\
    \end{tabular}
    \caption{Numerical details of DNS}
    \label{tab:DNS}
\end{table}

\subsection{Resolvent Analysis}
Our model solution will be based on the resolvent formulation of \citet{mckeon_critical-layer_2010}, which assumes that the one dimensional mean velocity profile $\overline{\mathbf{U}}(r)$ is known. This allows us to write (\ref{pNSE}) and (\ref{pcont}) as a balance between the linear dynamics and the nonlinear term which we group into a forcing term denoted by $\mathbf{f}$. 
\begin{equation}
     \mathcal{L}\mathbf{u}=\mathcal{N}(\mathbf{u},\mathbf{u}) \equiv \mathbf{f}.
\end{equation}
We Fourier transform both the velocity $\mathbf{u}(r,z,\theta,t)$ and the nonlinear forcing $\mathbf{f}(r,z,\theta,t)$ in time as well as the homogeneous spatial directions, $\theta$ and $z$. For a function $\boldsymbol{q}(r,z,\theta,t)$ the Fourier transform is defined as
\begin{equation} \label{FT}
    \hat{\boldsymbol{q}}(r,k,n,\omega)\equiv  \int_{-\infty}^{\infty}\int_{\infty}^{\infty}\int_{0}^{2\pi} \boldsymbol{q}(r,z,\theta,t)e^{-i(k_z z  +n\theta -\omega t) }\,d\theta\, dz\, dt.
\end{equation}
This results in system of coupled ordinary differential equations (ODEs) for the Fourier modes $\hat{\mathbf{u}}_{\mathbf{k}}(r)$ and $\hat{\mathbf{f}}_{\mathbf{k}}(r)$ parameterized by the wavenumber triplet $\mathbf{k}\equiv[k_z,n,\omega]$. The domain is periodic in the azimuthal direction and we formally consider the case $L_z = \infty$, so we have $n \in \mathbb{Z}$ and $k_z,\omega \in\mathbb{R}$.
\begin{equation}\label{resNSE}
    \left(\mathcal{L}_\mathbf{k}-i\omega M\right) \hat{\mathbf{u}}_{\mathbf{k}} =\hat{\mathbf{f}}_{\mathbf{k}}
\end{equation}
The explicit expressions for the Fourier transformed linear operator $\mathcal{L}_\mathbf{k}$ and the weight matrix $M$ are given in appendix \ref{appA}. If we then invert the linear operator on the left hand side of (\ref{resNSE}), we arrive at the characteristic input-output representation of the NSE of \citet{mckeon_critical-layer_2010} where the linear dynamics represent a transfer function from the nonlinearity, which is interpreted as a forcing, and the velocity.
\begin{equation}\label{resNSEi}
    \hat{\mathbf{u}}_{\mathbf{k}} = \left( \mathcal{L}_\mathbf{k} - i\omega M\right)^{-1}\hat{\mathbf{f}}_{\mathbf{k}}
\end{equation}
A singular value decomposition (SVD) of this transfer function which we call the resolvent and denote by $\mathcal{H}$ provides an orthonormal basis for the velocity as well as the forcing, where superscript $^H$ denotes the conjugate transpose.
\begin{equation}\label{H}
    \mathcal{H}_{\mathbf{k}} \equiv \left(\mathcal{L}_\mathbf{k} - i\omega M\right)^{-1} = \Psi_\mathbf{k} \Sigma_\mathbf{k} \Phi_\mathbf{k}^H
\end{equation}
The columns of $\Psi_\mathbf{k}$ and $\Phi_\mathbf{k}$ are denoted by $\boldsymbol{\psi}_{\mathbf{k},j}$ and $\boldsymbol{\phi}_{\mathbf{k},j}$ are referred to as the resolvent response and forcing modes respectively and are ordered by their singular values $\sigma_{\mathbf{k},j}$ such that $\sigma_{\mathbf{k},1}\geq\sigma_{\mathbf{k},2}\geq...\geq\sigma_{\mathbf{k},j}\geq\sigma_{\mathbf{k},j+1} $. Thus $\boldsymbol{\psi}_{\mathbf{k},1}$ represents the most linearly amplified structure at that wavenumber. Note that these modes are vector fields over $r$ which are orthonormal with respect to an $L_2$ inner product over the three velocity components

\begin{equation}\label{innerproduct}
    \langle \mathbf{a},\mathbf{b}\rangle \equiv \int_{r_i}^{r_o} a^*_m(r)b_m(r) \, r\, dr
\end{equation}
with associated norm 
\begin{equation}\label{innerproductnorm}
    \|\mathbf{a}\|\equiv \langle \mathbf{a},\mathbf{a}\rangle^{1/2}
\end{equation}
where summation over $m$ is implied, such that
\begin{equation}
    \langle \boldsymbol{\psi}_{\mathbf{k},i},\boldsymbol{\psi}_{\mathbf{k},j} \rangle = \langle \boldsymbol{\phi}_{\mathbf{k},i},\boldsymbol{\phi}_{\mathbf{k},j} \rangle = \delta_{ij}.
\end{equation}
In this basis, each Fourier mode of the velocity and forcing may be written as
\begin{equation}\label{uexp}
    \hat{\mathbf{u}}_\mathbf{k} = \sum_{j=1}^{\infty} \sigma_{\mathbf{k},j} \chi_{\mathbf{k},j} \boldsymbol{\psi}_{\mathbf{k},j}
\end{equation}
\begin{equation}\label{fexp}
    \hat{\mathbf{f}}_\mathbf{k} = \sum_{j=1}^{\infty}  \chi_{\mathbf{k},j} \boldsymbol{\phi}_{\mathbf{k},j}
\end{equation}
where $\chi_{\mathbf{k},j} \equiv \langle  \phi_{\mathbf{k},j} ,\hat{\mathbf{f}}_{\mathbf{k}} \rangle$ represents the projection of the (unknown) forcing onto the forcing modes.

\subsection{Symmetries of Taylor vortex flow}
The various flow states observed in TCF (TVF, WVF, and MWVF) may be defined by their spatio-temporal symmetries \citep{rand_dynamics_1982}. We define TVF, the focus of this study, as a solution to (\ref{NSE}) which is steady, axisymmetric, and axially periodic with fundamental wavenumber $\beta_z$, meaning we restrict ourselves to wavenumber vectors of the form $\mathbf{k}=[k\beta_z,0,0]$ where $k\in \mathbb{Z}$.  This fundamental wavenumber $\beta_z$ is related to the axial height of the Taylor vortices and is generally constrained by the experimental apparatus or computational box since the domain must contain an integer number of vortices. The resolvent formulation assumes an infinite axial domain so the choice of $\beta_z$ is not immediately obvious. However, we found that the results shown in this work are robust to changes in $\beta_z$ as long as $ \pi/2 \lesssim \beta_z \lesssim 4\pi/3$. Therefore, we choose the axial periodicity of our model to match that observed in our DNS, allowing for a direct comparison between our model and the DNS. The specific values of $\beta_z$ are listed in table \ref{tab:error}. Given these symmetries our model solution will consist of an expansion in Fourier modes
\begin{equation}\label{uexp2}
    \mathbf{u}(r,z) = \sum_{k=1}^{N_k} \hat{\mathbf{u}}_k(r)e^{i k\beta_z z}
\end{equation}
where each Fourier mode $\hat{\mathbf{u}}_k$ is itself an expansion in resolvent modes given by (\ref{uexp}). We truncate the model at $N_k$ Fourier modes each of which is expanded in $N^{k}_{SVD}$ resolvent modes such that the final form of the TVF solution is given by
\begin{equation}\label{uexp3}
    \mathbf{u}(r,z) = \sum_{k=1}^{N_k}\sum_{j=1}^{N^k_{SVD}} \sigma_{k,j} \chi_{k,j} \boldsymbol{\psi}_{k,j}(r)e^{i k\beta_z  z}.
\end{equation}

\subsection{Treatment of the nonlinearity}
At any given wavenumber, the forcing $\hat{\mathbf{f}}_{k}$ is given by a convolution of the interactions of all triadically compatible velocity modes. 

\begin{equation}\label{fconv}
    \hat{\mathbf{f}}_k= \sum_{k'\neq \mathbf{0}} -\nabla\cdot\left(\hat{\mathbf{u}}_{k'} \hat{\mathbf{u}}_{k-k'}\right) 
\end{equation}
This means that the forcing at a given wave number $k$ contains only interactions between Fourier modes whose wavenumbers add up to $k$. Throughout this work we use the terminology \textit{``$k_1 = k_2 + k_3$"} to refer to a single (resonant) triad involving the nonlinear interaction between Fourier modes with wavenumbers $k_2$ and $k_3$ forcing the Fourier mode with wavenumber $k_1$. 

Equating the two expressions for the forcing mode given by  (\ref{fexp}) and (\ref{fconv}) and substituting (\ref{uexp}) for the velocity modes gives
\begin{equation}\label{nonlin1}
    \sum_{j=1}^{\infty}  \chi_{k,j} \boldsymbol{\phi}_{k,j} = \sum_{k'\neq \mathbf{0}} \sum_{p=1}^{\infty} \sum_{q=1}^{\infty}- \sigma_{k',p} \chi_{k',p} \sigma_{k-k',q} \chi_{k-k',q} \nabla\cdot\left(   \boldsymbol{\psi}_{k',p}     \boldsymbol{\psi}_{k-k',q}\right) .
\end{equation}
Projecting both sides of  (\ref{nonlin1}) onto each forcing $\boldsymbol{\phi}_{k,i}$ and dropping the summation symbols for simplicity gives 
\begin{equation}\label{chiN}
      \chi_{k,i} =  \chi_{k',p}  \chi_{k-k',q} N_{kk',ipq} 
\end{equation}
where the $N_{kk',ipq}$ are called the interaction coefficients and are given by 
\begin{equation}\label{Ndef}
      N_{kk',ipq} = - \sigma_{k',p} \sigma_{k-k',q}  \langle \boldsymbol{\phi}_{k,i},\nabla\cdot\left( \boldsymbol{\psi}_{k',p}     \boldsymbol{\psi}_{k-k',q}\right)\rangle
\end{equation}
which, critically, can be computed solely from knowledge of the linear operator $\mathcal{H}$. 

Nonlinear interactions between the velocity fluctuations also appear in the divergence of the Reynolds stress on the right hand side of (\ref{RANS}). This term is referred to as the ``mean forcing" and is given by the sum of nonlinear interactions of all the $\hat{\mathbf{u}}_k$ and their complex conjugates  $\hat{\mathbf{u}}_{-k}$, which can be directly interpreted as (\ref{fconv}) evaluated at $k = \mathbf{0}$:
\begin{equation} \label{meanf}
      -\overline{\mathbf{u} \cdot\nabla \mathbf{u}} =  \sum_{k'\neq \mathbf{0}} -\nabla\cdot\left(\hat{\mathbf{u}}_{k'} \hat{\mathbf{u}}_{-k'}\right) \equiv\hat{\mathbf{f}}_\mathbf{0}. 
\end{equation}
At this point we would like to reiterate that the mean velocity profile is assumed to be known a priori. Thus the left hand side of (\ref{RANS}), and therefore the Reynolds stress divergence on left hand side of (\ref{meanf}), is also known. 

We have thus reduced the NSE (under the assumption of a known mean velocity) to the infinite system of coupled polynomial equations (\ref{chiN}) for the complex coefficients $\chi_{\mathbf{k},j}$ with the auxiliary condition that (\ref{meanf}) is satisfied. While deriving an exact (nontrivial) solution to  (\ref{nonlin1}) may be a daunting task, we will demonstrate that approximate solutions can be efficiently computed by minimizing the residuals associated with (\ref{chiN}) and (\ref{meanf}).

\subsection{Optimization Problem} \label{sec:optimization}
We have recast the NSE in the language of resolvent analysis as
\begin{equation}\label{resx}
      \chi_{k,j} -  \chi_{k',m}  \chi_{k-k',n} N_{kk',jmn} = 0,~\ \forall \, k,j
\end{equation}
\begin{equation} \label{res0}
      \overline{\mathbf{u} \cdot\nabla \mathbf{u}} - \mathbf{f}_{0,k',m n} \chi_{k',m} \chi^*_{k',n}  = \mathbf{0},
\end{equation}
\begin{equation}\label{deff0}
  \mathbf{f}_{0,k',mn} \equiv \sigma_{k',m}  \sigma_{k',n} \nabla\cdot\left(\hat{\boldsymbol{\psi}}_{k',m} \hat{\boldsymbol{\psi}}^*_{k',n}\right),
\end{equation}
where we have expanded the velocity Fourier modes in their resolvent basis according to (\ref{uexp}), and summation over $m, n$ and $k'$ is implied. We truncate the expansion at some number of harmonics, $N_k$, of the fundamental wavenumber, and at each retained harmonic we truncate the singular mode expansion at $N^k_{SVD}$ such that the total number of retained modes is $N=\sum^{N_k}_{k=1}N^k_{SVD}$.  We seek to minimize the residuals (in the sense of the $L_2$ norm) associated with (\ref{resx}) and (\ref{res0}), and thus formulate the following optimization problem:
\begin{equation}\label{optprob1}
    \min_{\chi_{k,j}} g^2(\chi_{k,j})  = a g^2_0(\chi_{k,j}) + (1-a) g^2_{triad}(\chi_{k,j}).
\end{equation}
The first and second terms on the right hand side in (\ref{optprob1}) are defined as the mean constraint,
\begin{equation}\label{mean con}
    g_0(\chi_{k,j}) \equiv \frac{\|\overline{\mathbf{u} \cdot\nabla \mathbf{u}} -\mathbf{f}_{0,k',m n} \chi_{k',m} \chi^*_{k',n}\|}{\|\overline{\mathbf{u} \cdot\nabla \mathbf{u}}\|},
\end{equation}
and the triadic constraint,
\begin{equation}\label{triad con}
    g_{triad}(\chi_{k,j}) \equiv |\chi_{k,j} -  \chi_{k',m}  \chi_{k-k',n} N_{kk',jmn}|.
\end{equation}
The former represents the residual in the mean momentum equation (\ref{RANS}), while the latter represents the residual in the equation for the fluctuations (\ref{pNSE}). The user defined weighting parameter $a\in(0,1)$ determines the relative penalization of each of these two constraints in the residual. 

At this point we would like to highlight several important aspects of problem (\ref{optprob1}). First, we reiterate that the left hand term in the mean constraint (\ref{mean con}) is a known function since the mean velocity profile is assumed to be known a priori. Second, we emphasize that we have assumed no closure model and made no modeling assumptions regarding the form of the nonlinear forcing in the derivation of (\ref{optprob1}). Finally, while in general the amplitudes $\chi_{k,j} \in \mathbb{C}$, for the special case of steady axisymmetric solutions considered here it can be shown that $\chi_{k,j} \in \mathbb{R} ~\ \forall ~\ j,k$ meaning the optimization need only be carried out over a real valued domain.

Finally, we note that while the reformulation of the NSE (\ref{pNSE}) in the resolvent framework (\ref{resx}) is reminiscent of a Galerkin method (GM) where the governing equations are projected onto some predetermined set of basis functions, the current approach is appreciably different. Since we consider a steady process we can not integrate the equations forward in time as would be generally done in a GM. Furthermore, since the resolvent framework provides a basis for both the velocity and the nonlinearity, (\ref{resx}) is an exact representation of (\ref{pNSE}) whereas a GM represents the governing equations only in an integral sense. 

\subsection{Solution Methodology}
We solve the optimization problem (\ref{optprob1}) using a trust-region algorithm built into Matlab's $fminunc$ function for $R = 100,~\ 200,$ and $ 400$. The gradient and Hessian of (\ref{optprob1}) may be derived explicitly and are input to the algorithm to improve accuracy.  The weighting parameter $a$ in (\ref{optprob1}) is set to 0.01 which means that the triadic constraint is penalized 99 times more heavily than the mean constraint. This reflects the observation that the triadic constraint which encodes the fully nonlinear governing equation for the fluctuations is far more complex than the mean constraint which, given the fact that the mean profile is known, is simply a least squares fit to a curve. 
 
While this value of $a$ was found to lead to the most consistent and accurate results, the results are qualitatively robust to changes in $a$ as long as $0.0005\lesssim a \lesssim 0.8$. If $a\lesssim0.0005$, i.e. the triadic constraint is weighted too heavily, the optimization converges to the laminar state since the triadic constraint admits a trivial solution. If $a\gtrsim0.8$, the mean constraint is weighted too heavily and the optimization tends to over fit to the input mean and converge to a local minimum which is not self-sustaining and inaccurately reconstructs the velocity field. 
 
The optimization also requires an initial guess. For this we solve the rank 1 formulation of (\ref{optprob1}) using just one wavenumber, the fundamental, and one resolvent mode, in which case the minimum can be found analytically, resulting in an amplitude $\chi_{rank 1} \approx 0.13$. We then initialize the full optimization such that $\chi_1 = \chi_{rank 1}$ and the remaining $\chi_{j>1}$ are assigned random values between -0.01 and 0.01. 

We assess the convergence and accuracy of our model solution using three metrics. First, we compute the final minimum residual of the cost function in (\ref{optprob1}) denoted $g^*$. Second, we compute the error of the model solution compared to the temporal average of the DNS solution

\begin{equation}\label{dns error}
    e_{dns} \equiv \sum_{k=1}^{N_k} \|\boldsymbol{\hat{u}}_k-\boldsymbol{\hat{u}}_{k,dns}\|.
\end{equation}
The norm is defined in (\ref{innerproductnorm}). Third we quantify the error of our model solution in solving the underlying governing equations (\ref{pNSE}), the details of which are discussed in \S\ref{sec:nonlinear}.

We note that the error metric comparing our model to the DNS should be viewed with some caution since the Fourier decomposition of the DNS involves some inherent uncertainty. While our model is formulated in Fourier space, the DNS to which we compare our model utilizes a finite difference method in the axial direction. This means that the five or six Taylor vortices in the computational domain are not necessarily exactly the same size and the fundamental wavenumber $\beta_z$ can only be defined in an average sense, 
\begin{equation}
    \beta_z = 2\pi \frac{n_{roll}}{L_z}
\end{equation}
where $n_{roll}$ is the number of Taylor vortices contained in the domain and $L_z$ is the axial domain size. We use this average $\beta_z$ in the construction of our model. To compute the DNS Fourier modes used in (\ref{dns error}) 
we extract a single Taylor vortex whose size is closest to the average and perform a Fourier decomposition on this reduced domain. The largest difference between this best fit wavenumber and the average we observed was $0.5\%$. Since the radial shape of the Taylor vortices is expected to differ slightly with axial size it is unclear whether minor differences between our results and the DNS
are due to errors in our model or uncertainties in the Fourier decomposition of the DNS.  

At $R=400$, which will be the main focus in this work, it was found that  $N_k=9$ axial wavenumbers with $N^k_{SVD} = 22$ for  $k\leq 4$ and $N^k_{SVD} = 10$ for $k>4$ resolvent modes were sufficient such that we did not observe any further meaningful decrease in the residual $g^*$ with increased $N$. A detailed discussion of the choice of these particular truncation values is presented in \S\ref{sec:model reduction}. These truncation values, the total degrees of freedom, the axial wavenumber $\beta_z$, as well as the error metrics for all three Reynolds numbers are summarized in table \ref{tab:error}. 
 
\begin{table}
    \centering
    \begin{tabular}{c|c|c|c}
        $R$ & 100 & 200 & 400\\
        \hline
        $\beta_z$ & 3.67 & 3.67 & 2.62\\
        $N_k$ & 4 & 8 & 9 \\
        $N_{DNS}^k$ & $12 ~\ \forall ~\ k$ & $12(k\leq4),~\ 8(k>4)$ & $22(k\leq4),~\ 10(k>4)$\\
        $N$ & 48 & 80 & 138 \\
        $g^*$ & $2.3\times10^{-3}$ & $3.5\times10^{-3}$ & $5.5\times10^{-3}$ \\
        $e_{dns}$ & $2.5\times10^{-3}$ & $4.8\times10^{-3}$ & $1.5\times10^{-2}$ \\
        $e_{0}$ & $1.8\times10^{-3}$ & $1.3\times10^{-3}$ & $6.0\times10^{-3}$ \\
        $e_{nse}$ & $1.2\times10^{-3}$ & $2.7\times10^{-3}$ & $1.5\times10^{-3}$ \\
    \end{tabular}
    \caption{Reynolds numbers, fundamental wavenumber $\beta_z$, truncation values,  degrees of freedom: $N$, final residuals (normalized by initial residual): $g^*$, and error metrics for the three model solutions presented.}
    \label{tab:error}
\end{table}

\section{Results}\label{sec:results}

\subsection{Velocity field reconstruction}
\begin{figure}
\begin{subfigure}{0.5\textwidth}
\centering
\includegraphics[trim = 190 0 40 0,scale=0.45]{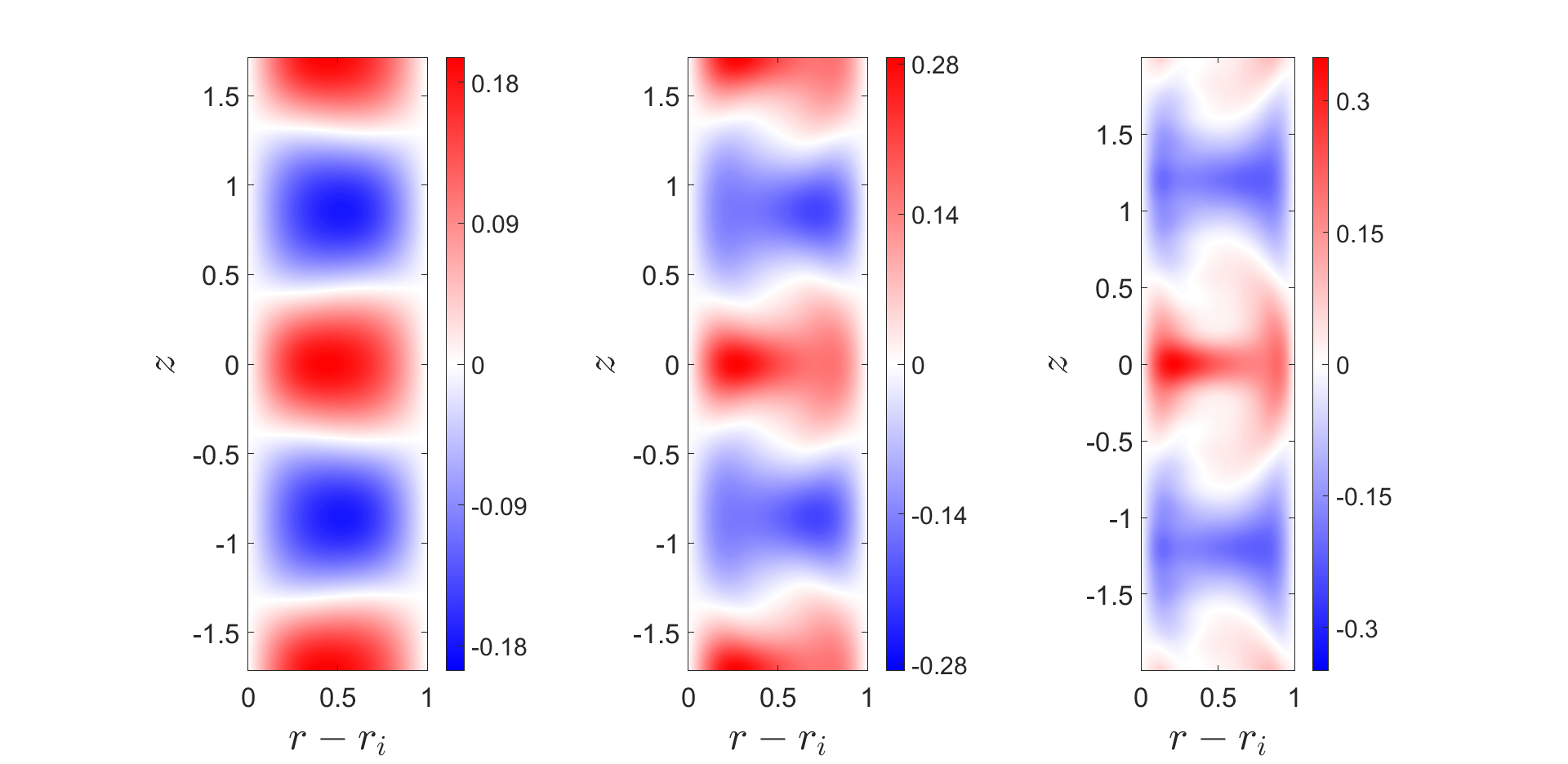} 
\end{subfigure}

\begin{subfigure}{0.5\textwidth}
\centering
\includegraphics[trim = 190 0 20 0,scale=0.45]{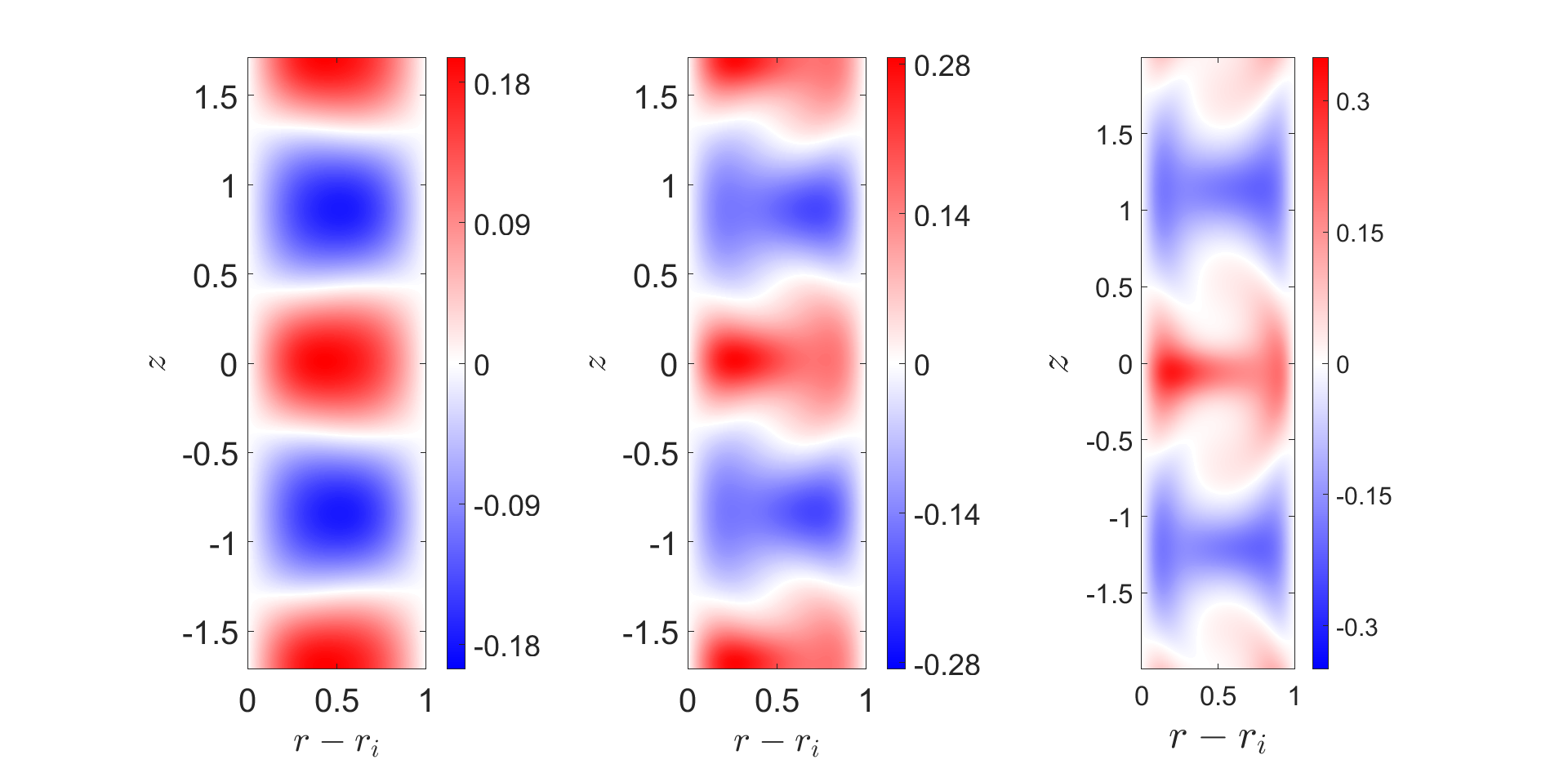} 
\end{subfigure}

\caption{Mean subtracted azimuthal velocity $u_{\theta}$ computed from our model (top row) and DNS (bottom row) at (from left to right) $R=100, ~\ 200$, and $400$.}
\label{fig:2d result u}
\end{figure}

\begin{figure}

\begin{subfigure}{0.5\textwidth}
\centering
\includegraphics[trim = 190 0 40 0,scale=0.45]{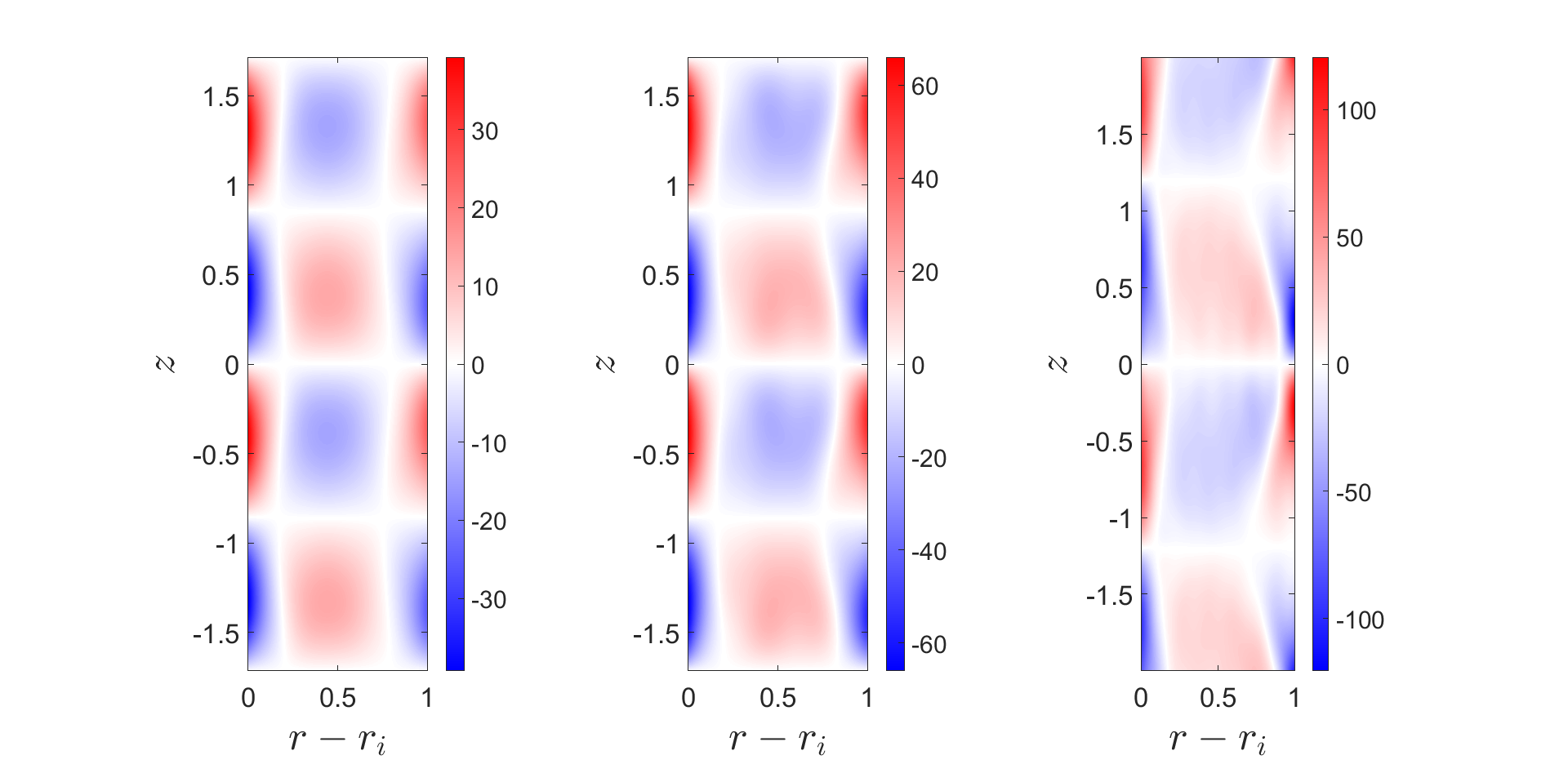} 
\end{subfigure}

\begin{subfigure}{0.5\textwidth}
\centering
\includegraphics[trim = 190 0 20 0,scale=0.45]{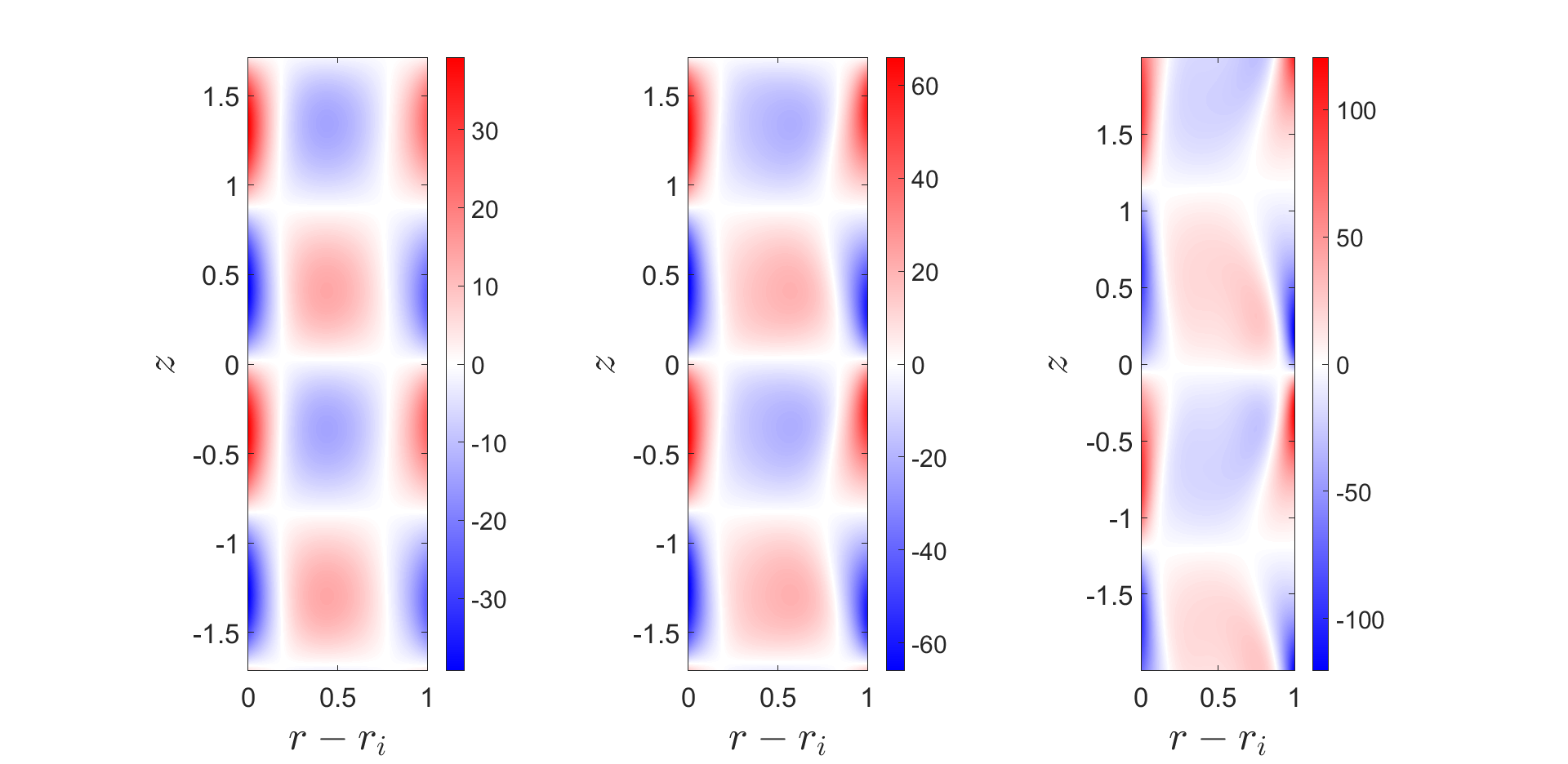} 
\end{subfigure}

    \caption{Azimuthal vorticity $\omega_{\theta}=\frac{\partial u_r}{\partial z} - \frac{\partial u_z}{\partial r}$ computed from our model (top row) and DNS (bottom row) at (from left to right) $R=100, ~\ 200$, and $400$.}
    \label{fig:2d result vort}
\end{figure}

The final result of the model is shown in figures \ref{fig:2d result u} and \ref{fig:2d result vort} where we compare the model result and the DNS. We plot the mean-subtracted azimuthal velocity $u_{\theta}$, and the azimuthal vorticity $\omega_{\theta}=\frac{\partial u_r}{\partial z} - \frac{\partial u_z}{\partial r}$ for $R=100, ~\ 200$ and $400$. The model solution is axisymmetric and steady by construction, and thus the radial and axial velocity are linked through continuity; no information is omitted by plotting $\omega_{\theta}$. As a comparison we show the azimuthal average of the mean subtracted DNS, however at this Reynolds number the flow is axisymmetric and steady, so the average field shown is representative of the flow at any azimuthal location and at any instance in time. There is good agreement between the resolvent model (top rows) and the DNS (bottom rows). The model accurately captures the dominant structure of the flow including the strong plumes of azimuthal velocity. The azimuthal vorticity exhibits a checkerboard pattern of regions of roughly constant vorticity of opposing signs. Regions of higher vorticity are concentrated near the walls, while the larger segments in the bulk of the domain have comparatively lower levels of vorticity. These results are in agreement with the DNS of \citet{sacco_dynamics_2019}, who found that as the Reynolds number increases this concentration of vorticity at the walls is enhanced and the bulk becomes increasingly ``empty" of vorticity.

\begin{figure}
\centering
\begin{subfigure}{0.45\textwidth}
\includegraphics[trim = 60 10 220 0,scale=0.45]{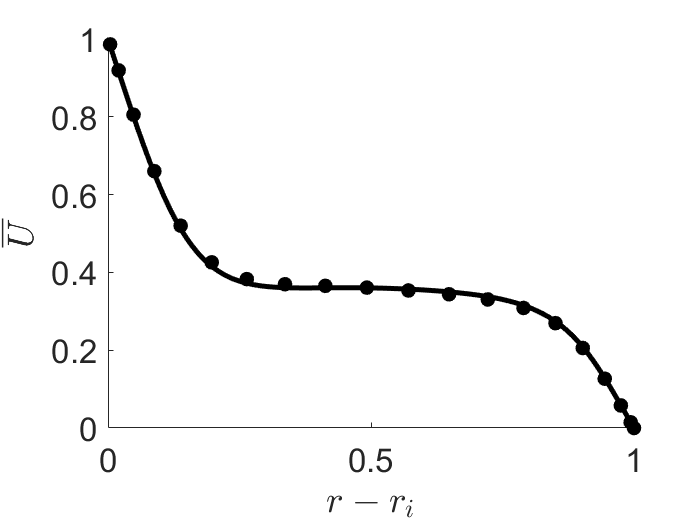}
\caption{}
\label{fig:result 400 U0}
\end{subfigure}
\begin{subfigure}{0.45\textwidth}
\includegraphics[trim = 0 10 220 0,scale=0.45]{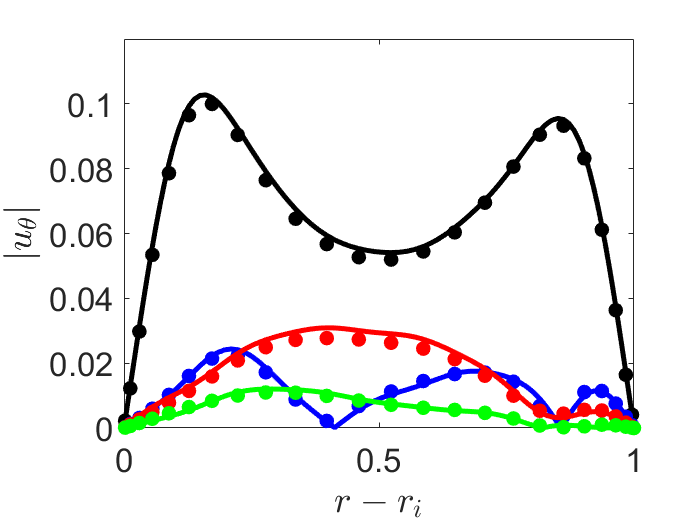}
\caption{}
\label{fig:result 400 ut}
\end{subfigure}
\centering
\begin{subfigure}{0.45\textwidth}
\includegraphics[trim = 60 10 220 0,scale=0.45]{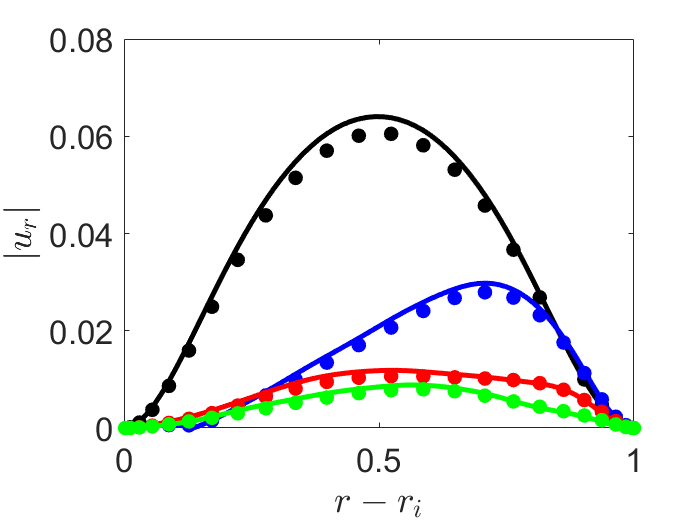}
\caption{}
\label{fig:result 400 ur}
\end{subfigure}
\centering
\begin{subfigure}{0.45\textwidth}
\includegraphics[trim = 0 10 220 0,scale=0.45]{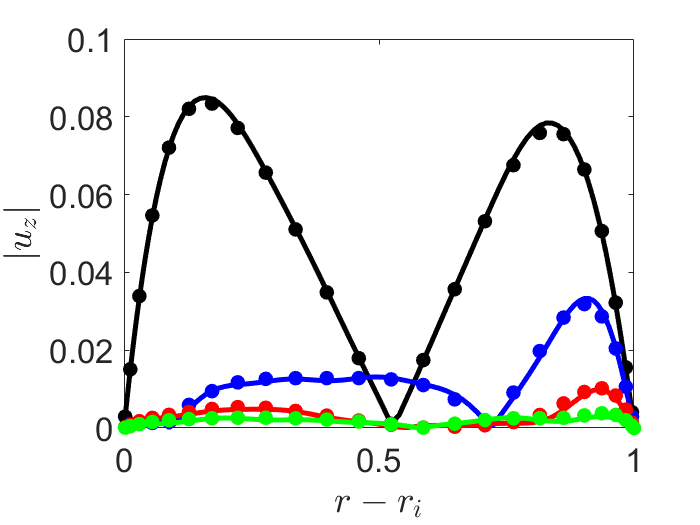}
\caption{}
\label{fig:result 400 uz}
\end{subfigure}
\caption{Model solution (lines) compared to the DNS (symbols) at $R = 400$. Mean velocity profile, $\bar{U}$, computed from Reynolds stress divergence of model compared to input mean velocity from DNS (\textit{a}). First five Fourier modes of model solution, $\hat{u}_{\theta}$, $\hat{u}_{r}$, $\hat{u}_{z}$ (\textit{b-d}), $k=1$ (black), $k=2$ (blue), $k=3$ (red), $k=4$ (green).}
\label{fig:1d result R400}
\end{figure}

A more quantitative assessment of the model's accuracy is shown in figure \ref{fig:1d result R400} where we compare the individual Fourier modes of the model solution to the Fourier modes computed from the DNS. For clarity of presentation we focus on $R=400$ and show only $(k\leq4)$, However an analysis of the accuracy of all retained Fourier modes for all Reynolds numbers is presented in figure \ref{fig:forcedmodes} in  \S\ref{sec:nonlinear}. Compared to the DNS, the model slightly over-predicts the amplitude of the radial velocity for the fundamental Fourier mode, but the wall parallel components of the fundamental are captured almost exactly. Most striking is the good agreement of the higher harmonics. The largest scale dominates the contribution to the Reynolds stress divergence and is thus determined primarily by the mean constraint, which as mentioned previously is relatively ``easy'' to solve. However the smaller scales require accurately approximating the solutions to the nonlinear triadic constraint, a much less trivial task. Furthermore, small deviations between the Fourier modes of the model and DNS are not necessarily indicative of errors in our model. For the reasons discussed above, a Fourier decomposition of the DNS incurs some inherent uncertainty. A more rigorous assessment of how accurately our model solves the governing equations is presented in \S\ref{sec:nonlinear}.  Additionally, figure \ref{fig:result 400 U0} compares the DNS mean velocity profile, used as an input to the model, to the mean velocity profile computed from Reynolds stress divergence of the model itself.  
\begin{equation}
    \frac{1}{R}\nabla^2\Bar{U}_{model} =  \sum^{N_k}_{=1} -\nabla\cdot\left(\hat{\mathbf{u}}_{k} \hat{\mathbf{u}}_{-k}\right)
\end{equation}
 
The input and output mean velocity profile show very good agreement, with only some mild discrepancy at the edge of the inner boundary layer. The error in mean velocity may be computed as
\begin{equation}\label{e_0}
    e_0 \equiv \|\Bar{U}_{DNS}-\Bar{U}_{model}\|
\end{equation}
which is associated with residual of the mean constraint (\ref{mean con}) in (\ref{optprob1}). However, note that while (\ref{e_0}) is written in terms of the mean velocity, (\ref{mean con}) is written in terms of the Reynolds stress divergence. The values of $e_0$ for all three Reynolds numbers are tabulated in table \ref{tab:error}. We generally do not use (\ref{e_0}) as one of the measures of convergence since very few modes are required to accurately capture the mean, and thus $e_0$ reaches a minimum long before the full nonlinear flow is converged. This is consistent with past studies which have have shown that the mean velocity profile of various flows may be accurately modeled using the Reynolds stress divergence of a single resolvent or eigen mode \citep{mantic-lugo_self-consistent_2014,mantic-lugo_self-consistent_2015,rosenberg_computing_2019}.

Overall, the success of the model in capturing this fully developed TVF indicates that, despite its fully nonlinear nature, the full solution remains relatively low dimensional. Note that our model constitutes approximately a million-fold reduction in degrees of freedom compared to the DNS. Nevertheless, given the relative simplicity of the flow, the model reduction is not as drastic as one might expect from an analysis purely of the energetic content of the flow. At $R=400$ the velocity associated with the third harmonic $(k=4)$ is two orders of magnitude less than the fundamental, and yet nine wavenumbers must be retained in order to achieve the convergence shown here. The dynamic importance of these energetically weak harmonics is discussed in \S\ref{sec:nonlinear}.


\subsection{Self sustaining solutions - closing the resolvent loop}\label{sec:nonlinear}
We have shown that our model accurately captures the structure of the TVF observed in the DNS. Now we analyze the accuracy of our model viewed from the perspective of a self sustaining process. In other words we assess how accurately our model approximates a solution to the governing equations (\ref{pNSE}). In the resolvent framework, the nonlinear term is interpreted as a forcing to the linear dynamics. As such, a solution is self-sustaining if the sum of all triadic interactions at a particular wavenumber provide the correct forcing for the response at that wavenumber. This means that we must have

\begin{equation}\label{selfsustain}
    \hat{\mathbf{u}}_k = \mathcal{H}_{k}\sum_{k'\neq \mathbf{0}} -\nabla\cdot\left(\hat{\mathbf{u}}_{k'} \hat{\mathbf{u}}_{k-k'}\right) ~\ \forall ~\ k.
\end{equation}
Note that (\ref{selfsustain}) is simply a restatement of (\ref{resNSEi}) with the nonlinear forcing written explicitly in terms of $\hat{\boldsymbol{u}}_{k}$ and that our model will generally not satisfy (\ref{selfsustain}) exactly.

Here we refer to the the direct result of our model, i.e. the quantity on the left hand side of (\ref{selfsustain}), as the  \textit{``primary"} velocity, and we denote right hand side of (\ref{selfsustain}), computed from that model solution, as the \textit{``forced"} velocity. In figure \ref{fig:forcedmodes} we plot the azimuthal component of both the primary and forced Fourier modes for all the wavenumbers and for all three Reynolds numbers.  For all Reynolds numbers and wavenumbers agreement between the primary (open circles) and forced mode (lines) is very good indicating that our model is indeed a close approximation of a solution to (\ref{pNSE}). Figure \ref{fig:forcedmodes} also shows the Fourier modes computed from the DNS for comparison (open squares). We see that there is growing discrepancy between the model result and the DNS modes with increasing wavenumber. However note that the discrepancy between the model and the DNS, which is only present in the higher harmonics, is two orders of magnitude smaller than the amplitude of the fundamental. This discrepancy is due to the structure of the nonlinear forcing and is discussed further in \S\ref{sec:nonlinear2}. Additionally, we note that for both the model and DNS the mode shapes of the higher harmonics, $k\gtrsim 3$ do not seem to differ significantly with increasing $k$ indicating some level of universality as the length scale decreases. 

We quantify the total error in nonlinear compatibility as
\begin{equation}\label{e_nse}
    e_{nse} = \sum_{k=1}^{N_k}\|\hat{\mathbf{u}}_k - \mathcal{H}_{k}\sum_{k'\neq \mathbf{0}} -\nabla\cdot\left(\hat{\mathbf{u}}_{k'} \hat{\mathbf{u}}_{k-k'}\right)\|. 
\end{equation}
This may be thought of as the total residual associated with how accurately our model solution approximates a solution to (\ref{pNSE}), or in other words it represents the final residual associated with the triadic constraint (\ref{triad con}). For all three Reynolds numbers considered the error is $\mathcal{O}(10^{-3})$, with the exact values listed in table \ref{tab:error}.

\subsection{Analysis of the forcing structure}\label{sec:nonlinear2}
\begin{figure}
    \centering
    \includegraphics[trim = 265 20 220 20,scale=0.4]{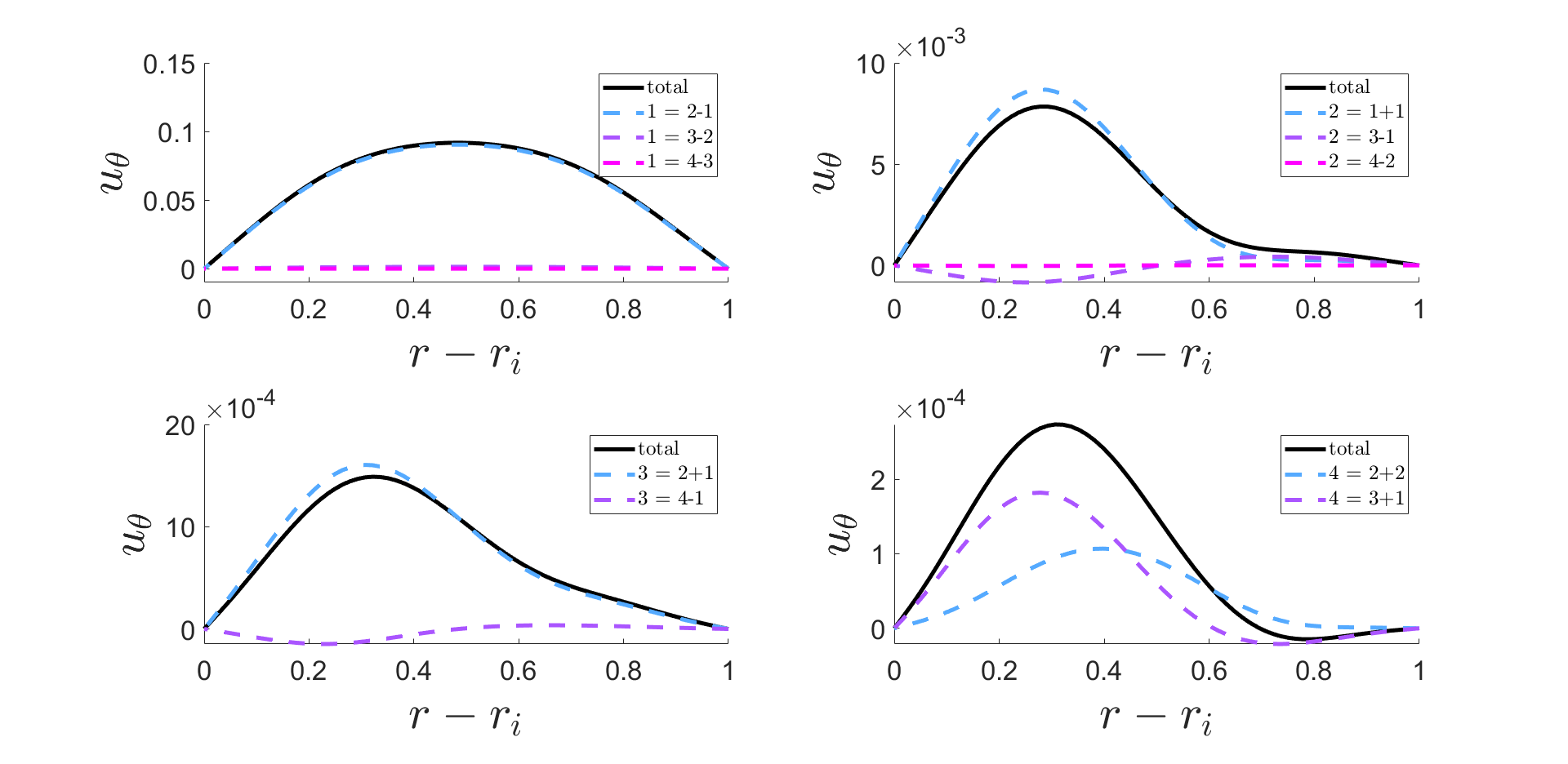}
    \caption{Azimuthal velocity component of the forced Fourier modes at $R=100$. The individual triadic contributions, $ \hat{\boldsymbol{v}}_{k,k'}$, are shown in dashed colors and the full Fourier mode, $\hat{u}_{k,\theta}$, is plotted in solid black. The sum of the individual triad components, (dashed lines) add up to the total forced mode (solid black).  The sum of the individual triad components, (dashed lines) add up to the total forced mode  (solid black). Top row $k=1-2$,  bottom row: $k=3-4$.  }
    \label{fig:forcedmodes2 100}
\end{figure}

\begin{figure}
    \centering
    \includegraphics[trim = 280 20 220 0,scale=0.25]{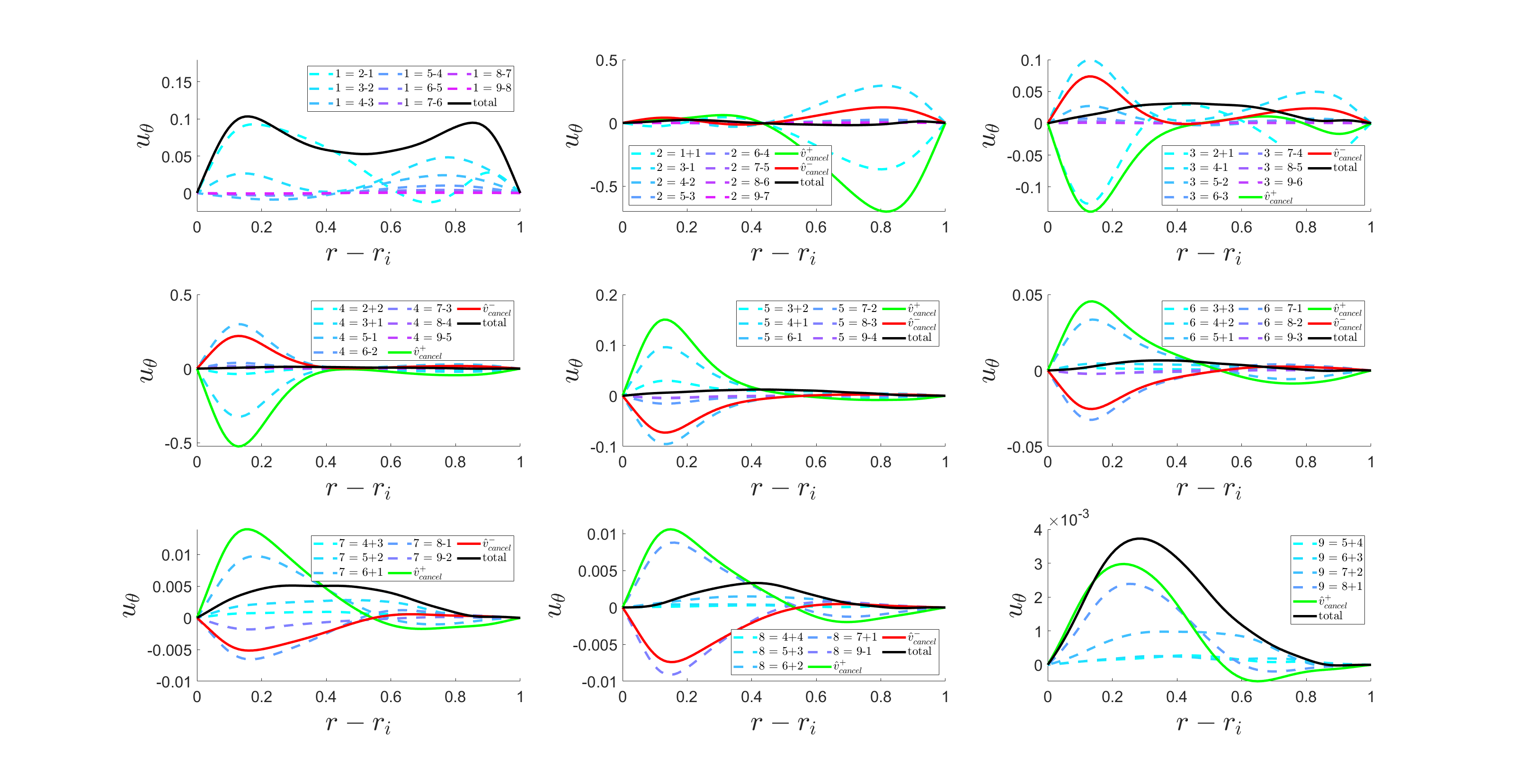}
    \caption{Azimuthal velocity component of the forced Fourier modes at $R=400$. The individual triadic contributions, $ \hat{\boldsymbol{v}}_{k,k'}$, are shown in dashed colors and the full Fourier mode, $\hat{u}_{k,\theta}$, is plotted in solid black. The sum of the individual triad components, (dashed lines) add up to the total forced mode (solid black). The predicted canceling azimuthal velocity contributions, $\hat{v}^{\pm}_{k,cancel,\theta}$, derived in \S\ref{sec:cancel} are plotted in red and green. Top row $k=1-3$, middle row $k=4-6$, bottom row: $k=7-9$. }
    \label{fig:forcedmodes2 400}
\end{figure}

Here we investigate which individual triadic interactions are most important in sustaining the flow and how these vary with Reynolds number. As previously noted, for $R=200,400$, the higher harmonics: $k\gtrsim 5$, do not contribute significantly to the energy content of the flow. Nevertheless, they play a crucial role in the nonlinear forcing of the larger structures and are necessary to achieve the convergence shown in figures \ref{fig:2d result u} and \ref{fig:2d result vort}. To identify the mechanics underlying the forcing structure we compute the individual terms in the summation on the right hand side of (\ref{selfsustain}). These individual contributions of the forced velocity, defined as
\begin{equation}\label{triad v contribution}
    \hat{\boldsymbol{v}}_{k,k'} \equiv -\mathcal{H}_{k}\nabla\cdot\left(\hat{\mathbf{u}}_{k'} \hat{\mathbf{u}}_{k-k'}\right),
\end{equation}
represent the contribution of each individual triadic interaction in (\ref{selfsustain}) and are shown in figures \ref{fig:forcedmodes2 100} and \ref{fig:forcedmodes2 400} for $R=100$ and $R=400$ respectively. The individual contributions (\ref{triad v contribution}) are plotted in dashed colored lines and the full Fourier mode is plotted in solid black. By definition the individual contributions (dashed colors) sum to the full Fourier mode (solid black). To clarify the following discussion we define a \textit{``(forward) forcing cascade"} as the forcing of mode $k_0$ by interactions involving strictly modes $k \leq k_0$ and an \textit{``inverse forcing cascade"} as the forcing of mode $k_0$ by interactions involving $k > k_0$.

At $R=100 \approx 1.25 R_c$, we observe a forcing mechanism reminiscent of a weakly nonlinear theory where the harmonics are all driven exclusively by a forward forcing cascade. The $k=2$ mode is driven primarily by the self interaction of the $k=1$ mode, the $k=3$ mode is driven by the interaction of the $k=1$ and $k=2$ modes, and so on. In other words, modes with wavenumber $k_0$ do not contribute to the forcing of modes with wavenumber $k<k_0$.

\begin{figure}
    \centering
    \includegraphics[trim = 265 20 220 20,scale=0.4]{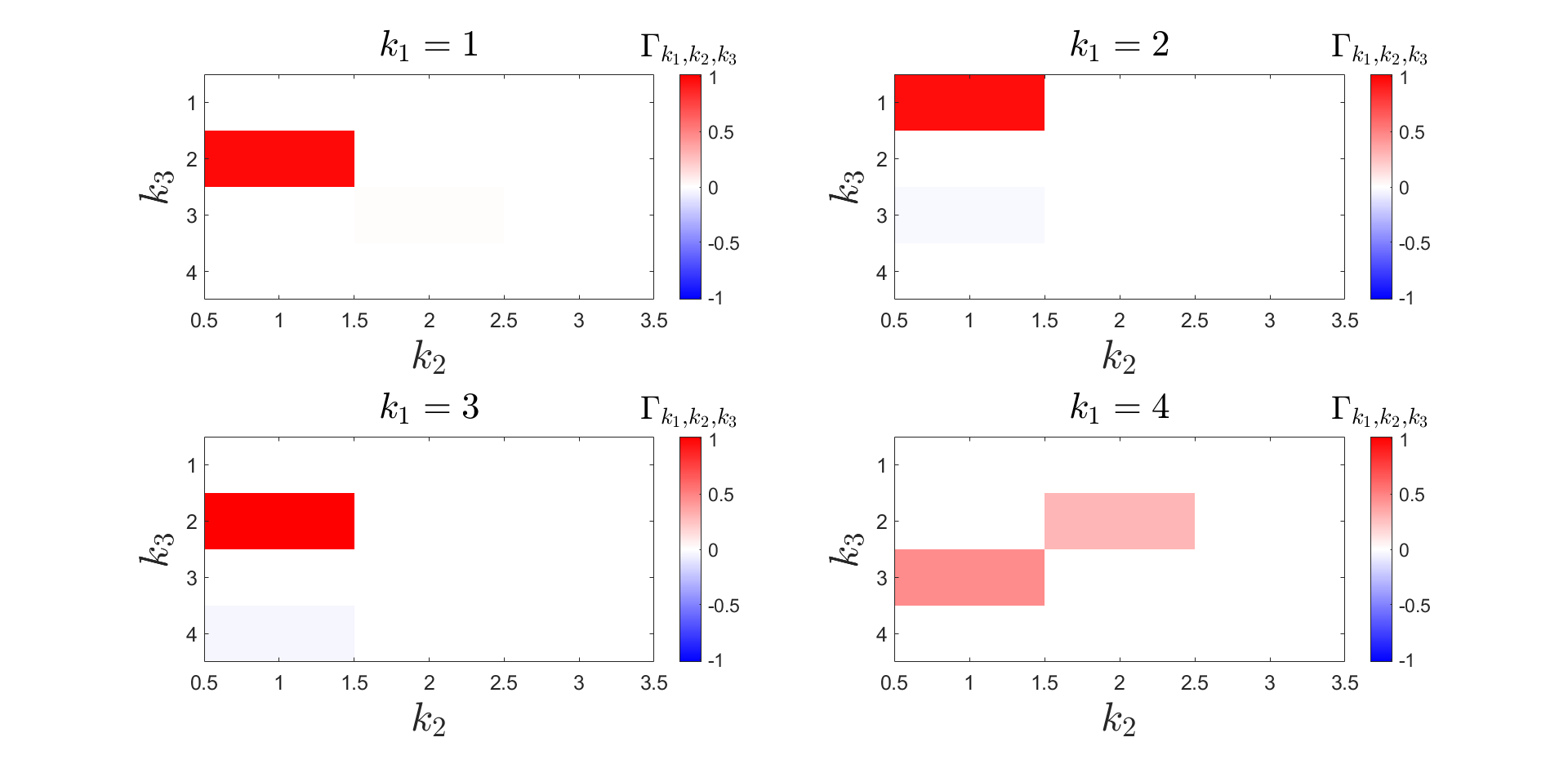}
    \caption{Projections of the velocity due to individual triadic interactions onto the full Fourier mode, $\Gamma_{k_1,k_2,k_3}$, at $R=100$. Top row $k=1-2$, bottom row: $k=3-4$. }
    \label{fig:triad projections 100}
\end{figure}

\begin{figure}
    \centering
    \includegraphics[trim = 200 95 0 5,scale=0.25]{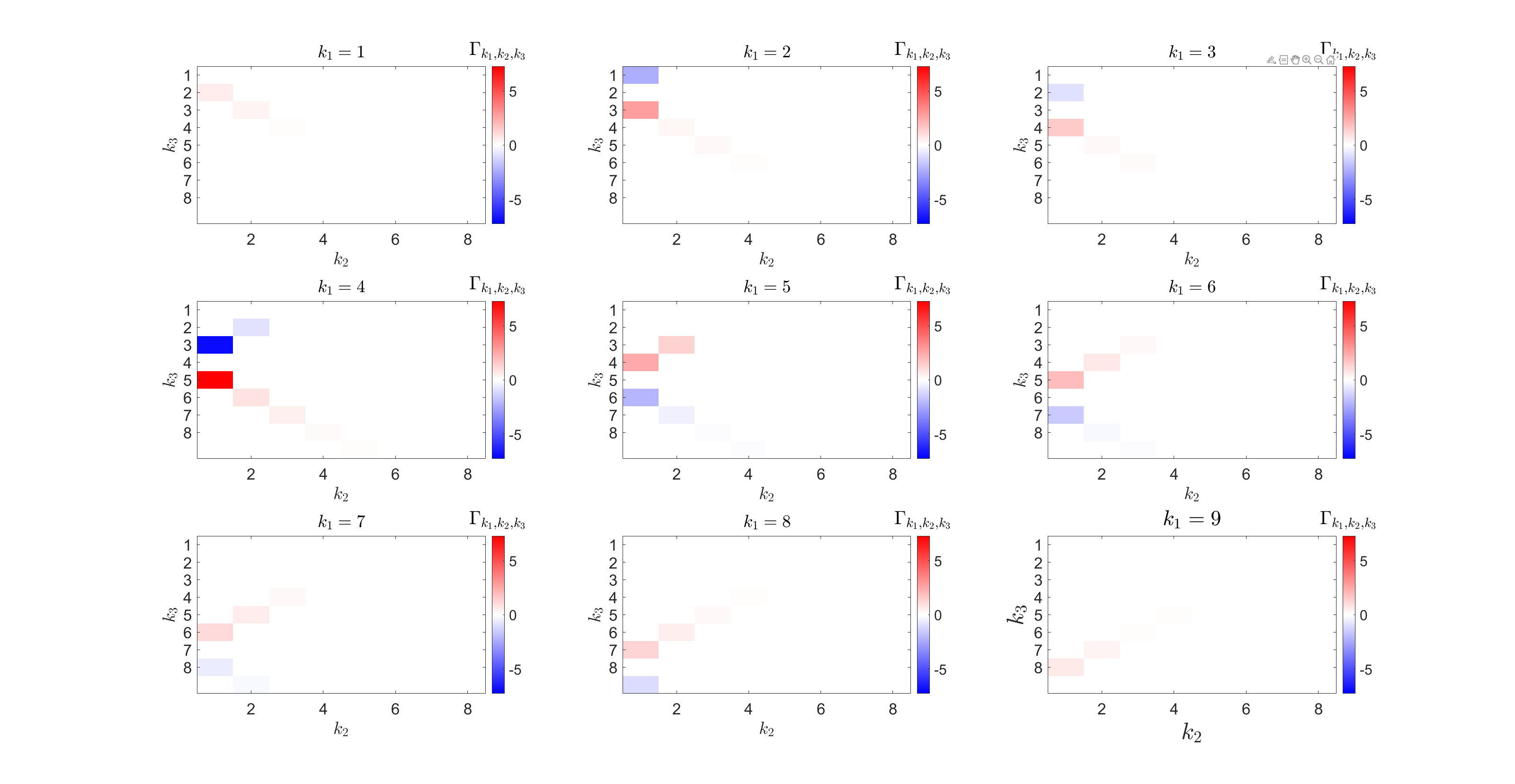}
    \caption{Projections of the velocity due to individual triadic interactions onto the full Fourier mode, $\Gamma_{k_1,k_2,k_3}$, at $R=400$. Top row $k=1-3$, middle row $k=4-6$, bottom row: $k=7-9$. }
    \label{fig:triad projections 400}
\end{figure}

\begin{figure}
    \centering
    \includegraphics[trim = 200 95 0 5,scale=0.25]{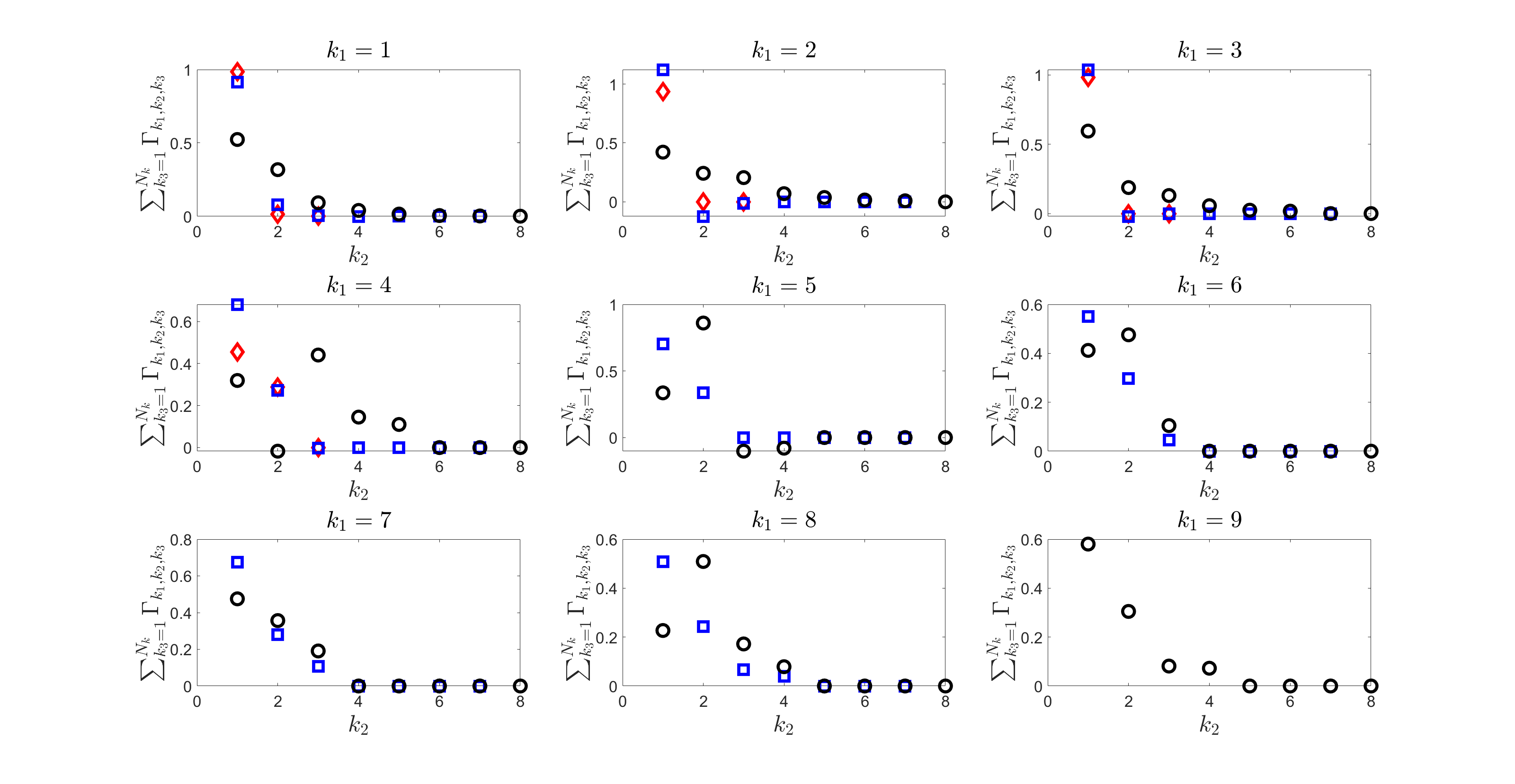}
    \caption{Projections of the velocity due to individual triadic interactions onto the full Fourier mode summed over common wavenumbers. $R=100$ in red, $R=200$ in blue squares, and $R=400$ in black circles. Top row $k=1-3$, middle row $k=4-6$, bottom row: $k=7-9$. }
    \label{fig:important triads}
\end{figure}

The higher Reynolds number model solutions do not exhibit the same unidirectional forcing cascade observed close to the bifurcation from laminar flow. We plot the same decomposition of 
(\ref{triad v contribution}) for $R=400\approx 5 R_c$ in figure \ref{fig:forcedmodes2 400}. Due to the significant disparity in amplitude between the individual triadic contributions and the composite mode we plot the two against separate y-axis. For the harmonics $(k>1)$, the pair of contributions due to triadic interactions involving the fundamental ($k=\pm1$), i.e. $k = (k+1) - 1$ and $k = (k-1) +1$, have large, almost equal amplitudes but are of opposite sign and almost cancel. The same phenomenon is observed for the triads involving the $k=\pm2$ mode, albeit it is not as pronounced as for the triads involving the fundamental. This raises the question of whether or not these components exactly cancel, and thus do not play a significant role in the dynamics, or whether the small differences in shape and amplitude dictate the structure of the resulting mode. To investigate this we compute the projection of the individual triadic contributions onto the full mode
\begin{equation}
    \Gamma_{k,k-k',k'} \equiv \frac{\langle \hat{\boldsymbol{v}}_{k,k'}, \hat{\mathbf{u}}_k\rangle}{\langle \hat{\mathbf{u}}_k, \hat{\mathbf{u}}_k\rangle}.
\end{equation}
These projections are plotted in figures \ref{fig:triad projections 100} and \ref{fig:triad projections 400} for $R=100$, and $R = 400$ respectively. Each sub-figure corresponds to one Fourier mode, $k_1$, and each tile represents the contribution to the $k_1$'th Fourier mode from the triadic interaction between $k_2$ and $k_3$. Note that the $k$'th column includes contributions from $\pm k$ since $\hat{\boldsymbol{u}}_{-k} = \hat{\boldsymbol{u}}_{k}^*$ and therefore both represent the same structure in physical space. As expected from figure \ref{fig:forcedmodes2 400}, we observe pairs of strong negative and positive correlations from the two triads involving the fundamental, with less pronounced, but still evident, pairing between triads involving $k = \pm2$.  In order to quantify the total contribution of these pairs of triads we sum the columns in figures \ref{fig:triad projections 100}, \ref{fig:triad projections 400}, and the equivalent case for $R=200$, the results of which are shown in figure \ref{fig:important triads}. This can be thought of as a measure of the importance of all triadic combinations involving a certain wavenumber. Note that by construction all the entries in each sub-figure of figures \ref{fig:triad projections 400} and \ref{fig:important triads} sum to one.  Figure \ref{fig:important triads} makes it clear that, for a given $k$, it is the two pairs of triads  $k = (k+1) - 1$ and $k = (k-1) +1$ as well as  $k = (k+2) - 2$ and $k = (k-2) +2$ which provide the dominant share of the forcing. We note that similar instances of destructive interference have been observed by other authors such as \citet{nogueira_forcing_2021} in their analysis of forcing statistics in plane Couette flow and \citet{rosenberg_efficient_2019} in their interpretation of the Orr-Sommerfeld/Squire decomposition of the resolvent operator. 

The large scales at $R=200$ are driven almost entirely by the pair of triads involving $k=\pm 1$ with the pair involving $k=\pm 2$ only becoming active for $k\geq4$. At $R = 400$ the forcing is more distributed among the various triadic interactions indicating a higher degree of nonlinearity. However, the triads involving $k=\pm1$ and $k=\pm2$ are clearly still dominant. In fact the contribution of the triads involving $k=\pm2$ is comparable and sometimes greater than the contributions from those involving $k=\pm1$. Nevertheless, it is the triads involving the fundamental which have the largest amplitude contributions and display the largest degree of destructive interference. 

\begin{figure}
    \centering
    \includegraphics[trim = 280 20 220 0,scale=0.25]{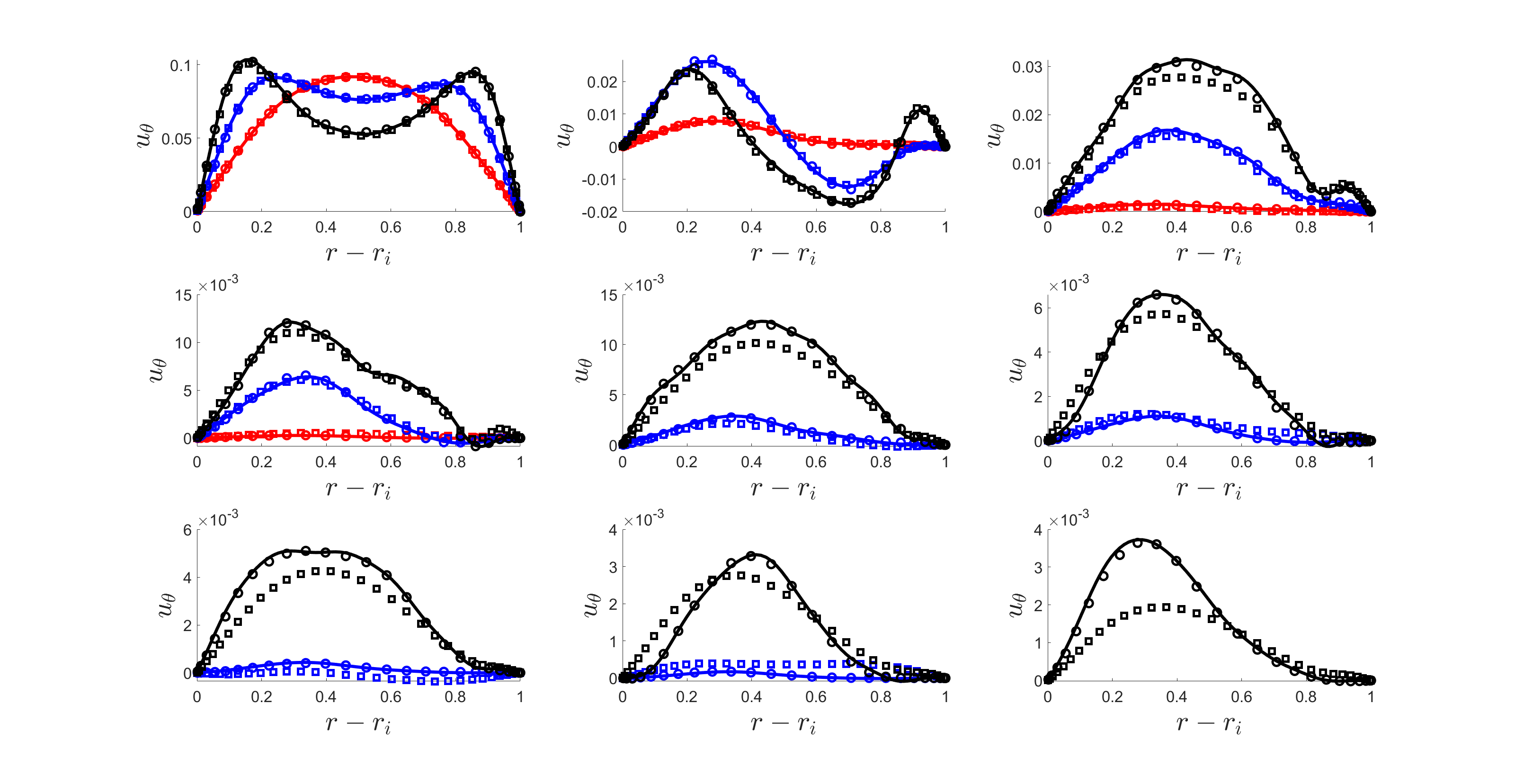}
    \caption{Azimuthal velocity component of the model solution's primary Fourier mode (open circles) and forced Fourier mode (lines) as well as the Fourier modes from DNS (open squares) for $R=100$ (red), $R=200$ (blue), and $R=400$ (black). Top row $k=1-3$, middle row $k=4-6$, bottom row: $k=7-9$. }
    \label{fig:forcedmodes}
\end{figure}

At this point we would like to revisit the discrepancy between the Fourier modes of the model solution and the DNS observed in figure \ref{fig:forcedmodes}. Recall that due to the structure of the forcing, an accurate reconstruction of a particular Fourier mode with wavenumber $k_0$ requires accurate knowledge of its harmonic, $k_0+1$. Practically, the model must be truncated at some point and so there will always be a maximum wavenumber $k_{m}$ whose harmonic $k_{m}+1$ is unknown. Therefore there will be some error in the reconstruction of $\hat{\boldsymbol{u}}_{k_{m}}$ since $\hat{\boldsymbol{u}}_{k_{m}+1}$ is unavailable to participate in the inverse forcing cascade described above. This error will then ``back propagate" through Fourier space until it is outweighed by the influence of the large scales and the constraint imposed on those large scales by the input mean flow. A detailed analysis of how the Fourier space truncation affects the convergence of the optimization is beyond the scope of this work, however it is interesting to note that, while the higher harmonics of our model solution deviate slightly from the DNS, they remain nonlinearly compatible to a very good approximation. In other words, the primary and forced Fourier modes of the model solution in figure \ref{fig:forcedmodes} agree very well as quantified by the small residuals as defined by (\ref{e_nse}) and listed in table \ref{tab:error}. 


\subsection{The transition from weakly to fully nonlinear Taylor vortices}
Many studies have approached the nonlinear modeling of TVF through weakly nonlinear (WNL) theory, where the general premise is that the structure of the largest scale is given by the critical eigenmode and that the higher harmonics are all derived from that fundamental mode \citep{stuart_non-linear_1960,yahata_slowly-varying_1977,jones_nonlinear_1981,gallaire_pushing_2016}. Despite being formally valid for only a small range of Reynolds numbers close to $R_c$, the mathematical difficulties associated with the nonlinearity of the NSE often necessitate the use of WNL methods outside this domain of validity \citep{gallaire_pushing_2016}. Our results illuminate the physical mechanisms which lead to the eventual failure of WNL theory as the Reynolds number increases. WNL theory proceeds by expanding the solution in an asymptotic series about the bifurcation point such that the leading order solution $u_0$ is the laminar base flow and the $\mathcal{O}(\epsilon)$ solution $u_1$ is given by the critical eigen mode. The $\mathcal{O}(\epsilon^2)$ solution $u_2$ as well as the mean flow correction is then found by solving the linear system forced by the nonlinear self interaction of $u_1$. The higher order terms may then be similarly computed sequentially by solving a forced linear system of the form $\mathcal{L}_k u_k = f(u_1,u_2,...,u_{k-1})$. At the lowest Reynolds number considered here, $R=100$, this formulation is valid since as shown in figures \ref{fig:forcedmodes2 100} and \ref{fig:triad projections 100}  the forcing for a certain $\hat{\boldsymbol{u}}_k$ depends only on interactions between larger scales. However, as discussed in \S\ref{sec:nonlinear2}, at higher Reynolds numbers the forcing is dominated by pairs of triads, one of which involves $\hat{\boldsymbol{u}}_{k+1}$, a mechanism which is impossible in the WNL formulation. 

This means that near the bifurcation from the laminar state a model solution of the nonlinear flow may be truncated at the highest wavenumber of interest since a given wavenumber depends only on its sub-harmonics. We define such a flow to be in the \textit{``weakly nonlinear"} (WNL) regime. As the Reynolds number increases the forcing cascade is no longer only from large to small scales and an equally important inverse cascade mechanism emerges. In this case we define the flow to be in the \textit{``fully nonlinear"} (FNL) regime. For the case of $\eta=0.714$ considered here, this transition occurs around some $100<R<200$. These findings indicate that if one desires to model a certain number of harmonics of a given flow, the expansion must be carried out to significantly higher order than the highest harmonic of interest.

 \citet{sacco_dynamics_2019} noted a similar transition in the dynamics of Taylor vortices. They note that while Taylor vortices first arise due to a supercritical centrifugal instability of the laminar base flow, for $R\sim\mathcal{O}(10^4)$ they persist in the limit of zero curvature i.e. in the absence of centrifugal effects \citep{nagata_three-dimensional_1990,sacco_dynamics_2019}. At sufficiently high Reynolds numbers, they find that the temporal evolution of the r.m.s. velocity associated with the Taylor vortex and the mean shear are perfectly out of phase and fluctuate with a common characteristic frequency. Their finding indicates a regenerative self sustaining process similar to the framework suggested by \citet{waleffe_self-sustaining_1997,hamilton_regeneration_1995}. Since we consider steady TVF it is difficult to make a direct comparison between their results and ours. However, it is possible that the transition from weakly to fully nonlinear Taylor vortices that we observe is the genesis of the type of self -sustaining Taylor vortices described by \citet{sacco_dynamics_2019}.

\subsection{Destructive interference forcing structure}\label{sec:cancel}
As described in \S\ref{sec:nonlinear}, we observe that in the FNL regime a crucial component of the forcing at a given wavenumber $k$ is the destructive interference of the two triads $k = (k-1) + 1$ and $k = (k+1) - 1$. This pair of triads lead to velocity contributions with large amplitudes but with opposite sign. This means that an accurate reconstruction of a Fourier mode with wavenumber $k$ requires knowledge of both its subharmonic $k-1$ and its harmonic $k+1$. In this section we show that for streamwise constant and spanwise periodic solutions, as considered in this work, this large amplitude destructive interference is a direct consequence of the structure of the Fourier representation of the nonlinear term $\boldsymbol{u}\cdot \nabla \boldsymbol{u}$. In cylindrical coordinates the nonlinear interaction between two axisymmetric Fourier modes $\boldsymbol{a} = [a_r, a_{\theta}, a_z]$ and $\boldsymbol{b} = [b_r, b_{\theta}, b_z]$ with axial wavenumbers $k_a$ and $k_b$ is given by
\begin{equation}
    \boldsymbol{f}_{a,b} \equiv \boldsymbol{a}\cdot\nabla \boldsymbol{b} + \boldsymbol{b}\cdot\nabla \boldsymbol{a}
\end{equation}
where the axial derivative in the gradient operator is replaced by multiplication by $i k_b$ and $i k_a$ respectively. For clarity of exposition we limit the following analysis to the azimuthal component of the forcing and note that analogous arguments hold for the remaining two components. The forcing at wavenumber $k$ due to the interactions of $k-1$ and $1$ is given by 
\begin{equation}\label{f_plus 1}
    \hat{\boldsymbol{f}}^{+} \equiv \hat{\boldsymbol{u}}_{1}\cdot\nabla \hat{\boldsymbol{u}}_{k-1} + \hat{\boldsymbol{u}}_{k-1}\cdot\nabla \hat{\boldsymbol{u}}_{1}.
\end{equation}
Using the continuity equation to eliminate the axial velocity, the azimuthal component takes the form
\begin{multline}\label{f_plus 2}
    f^{+}_{\theta} =  \left( \hat{u}_{1,r} \hat{u}'_{k-1,\theta} + \hat{u}_{k-1,r} \hat{u}'_{1,\theta} + \frac{\hat{u}_{1,\theta}\hat{u}_{k-1,_r}}{r}+ \frac{\hat{u}_{k-1,\theta}\hat{u}_{1,r}}{r} \right) - \\ (k-1)\frac{(r\hat{u}_{1,r})'\hat{u}_{k-1,\theta}}{r} - \frac{(r\hat{u}_{k-1,r})'\hat{u}_{1,\theta}}{r(k-1)}.
\end{multline}
Similarly,  forcing due to the interactions of $k+1$ and $-1$ is given by 
\begin{equation}\label{f_minus 1}
    \hat{\boldsymbol{f}}^{-} \equiv \hat{\boldsymbol{u}}_{-1}\cdot\nabla \hat{\boldsymbol{u}}_{k+1} + \hat{\boldsymbol{u}}_{k+1}\cdot\nabla \hat{\boldsymbol{u}}_{-1},
\end{equation}
with the azimuthal component taking the form
\begin{multline}\label{f_minus 2}
    f^{-}_{\theta} = \left( \hat{u}_{1,r} \hat{u}'_{k+1,\theta} + \hat{u}_{k+1,r} \hat{u}'_{1,\theta} + \frac{\hat{u}_{1,\theta}\hat{u}_{k+1,_r}}{r}+ \frac{\hat{u}_{k+1,\theta}\hat{u}_{1,r}}{r} \right) + \\ (k+1)\frac{(r\hat{u}_{1,r})'\hat{u}_{k+1,\theta}}{r} + \frac{(r\hat{u}_{k+1,r})'\hat{u}_{1,\theta}}{r(k+1)}.
\end{multline}
Here the superscript $'$ denotes partial derivatives with respect to $r$. Additionally, note that for the streamwise constant fluctuations considered here $\hat{\boldsymbol{u}}_{-1} = \hat{\boldsymbol{u}}_{1}^* = [\hat{u}_{1,r},\hat{u}_{1,\theta},-\hat{u}_{1,z}]$. 

For some integer wavenumber $k>1$, the Fourier modes associated with the nearest neighbor wavenumbers $k\pm1$ are defined as
\begin{equation}\label{Fukp1}
    \hat{\boldsymbol{u}}(r)_{k\pm1}\equiv \int_{-\infty}^{\infty}\boldsymbol{u}(r,z)e^{i\beta_z  (k\pm 1) z} dz = \int_{-\infty}^{\infty}\boldsymbol{u}(r,z)e^{i\beta_z k (1\pm \frac{1}{k}) z} dz.
\end{equation}
Since the destructive interference is most pronounced for small scales, we formally consider the case of $k \gg 1$, for which we can expand (\ref{Fukp1}) in a Taylor series about $k^{-1}=0$.
\begin{equation}
    \hat{\boldsymbol{u}}_{k\pm1} = \int_{-\infty}^{\infty}\boldsymbol{u}\left(e^{i\beta_z k z} \pm i \beta_z k^{-1} e^{i\beta_z k z} + \mathcal{O}(k^{-2})\right) dz = \hat{\boldsymbol{u}}_{k} + \mathcal{O}(k^{-1}).
\end{equation}
This observation that for $k\gg1$ $\hat{\boldsymbol{u}}_k$ and $\hat{\boldsymbol{u}}_{k\pm1}$ differ by a quantity which is $\mathcal{O}(k^{-1})$ meaning that for large values of $k$ the shape of the Fourier modes does not change drastically with increasing $k$. Figure \ref{fig:forcedmodes} shows that this is indeed the case. Substituting $\hat{\boldsymbol{u}}_{k\pm1} =  \hat{\boldsymbol{u}}_{k} + \mathcal{O}(k^{-1})$ into (\ref{f_plus 2}) and (\ref{f_minus 2}) we find at leading order 

\begin{equation}\label{f_plus 3}
    f^{+}_{\theta} = f_{\theta,eq} - k\frac{(r\hat{u}_{1,r})'\hat{u}_{k,\theta}}{r} + \mathcal{O}(k^{-1})
\end{equation}
\begin{equation}\label{f_minus 3}
    f^{-}_{\theta} = f_{\theta,eq} + k\frac{(r\hat{u}_{1,r})'\hat{u}_{k,\theta}}{r} + \mathcal{O}(k^{-1})
\end{equation}
where $f_{\theta,eq}$ is the same for both triads and is given by
\begin{equation}
    f_{\theta,eq} = \left[ \hat{u}_{1,r} \hat{u}'_{k,\theta} + \hat{u}_{k,r} \hat{u}'_{1,\theta} + \frac{\hat{u}_{1,\theta}\hat{u}_{k,_r}}{r}+ \frac{\hat{u}_{k,\theta}\hat{u}_{1,r}}{r} + \frac{(r\hat{u}_{1,r})'\hat{u}_{k,\theta}}{r} \right].
\end{equation}

The only remaining terms are equal in magnitude but of opposite sign. Furthermore, since both  $\hat{u}_{1,r}$ and $\hat{u}_{k,\theta}$ are bounded and nonzero these terms will scale proportionally with $k$ and therefore are expected to have large amplitudes since $k \gg 1$. Expressions (\ref{f_plus 3}) and (\ref{f_minus 3}) predict that the large amplitude destructive interference observed in figures \ref{fig:forcedmodes2 400} and \ref{fig:triad projections 400} occurs through the terms $\pm  k\frac{(r\hat{u}_{1,r})'\hat{u}_{k,\theta}}{r}$. 

Similar expressions can be derived for the other two coordinates such that the two vector forcing terms which are expected to cancel are given by
\begin{equation}\label{f_cancel}
    \hat{\boldsymbol{f}}^{\pm}_{k,cancel} \equiv \mp k\frac{(r\hat{u}_{1,r})'}{r}\hat{\boldsymbol{u}}_{k} \approx \mp k\frac{(r\hat{u}_{1,r})'}{r}\hat{\boldsymbol{u}}_{k\mp1} ~\ (k\gg 1),
\end{equation}
where the approximate equivalence on the right hand side is due to (\ref{Fukp1}).
This prediction may be tested by computing the corresponding velocity contributions to the Fourier mode with wavenumber $k$, given by
\begin{equation}
    \hat{\boldsymbol{v}}^{\pm}_{k,cancel} = \mathcal{H}_k \hat{\boldsymbol{f}}^{\pm}_{k,cancel}, ~\ (k\gg1),
\end{equation}
and comparing its shape to that of the total velocity contributions $\hat{\boldsymbol{v}}_{k,\pm 1}$ defined in (\ref{triad v contribution}). Figure \ref{fig:forcedmodes2 400} shows the $\hat{\boldsymbol{v}}^{\pm}_{k,cancel}$  alongside the individual  $\hat{\boldsymbol{v}}_{k,k'}$, and we see that the former is a quite accurate approximation of $\hat{\boldsymbol{v}}_{k,\pm1}$ for $k>3$. Surprisingly, we see that the prediction holds at least approximately even for $k = 2,3$ despite the derivation having assumed that $k \gg 1$. These findings establish that $\hat{\boldsymbol{f}}^{\pm}_{k,cancel}$ is indeed responsible for the large amplitude destructive interference characteristic of the fully nonlinear regime. 

\begin{figure}
    \centering
    \includegraphics[scale = 0.65]{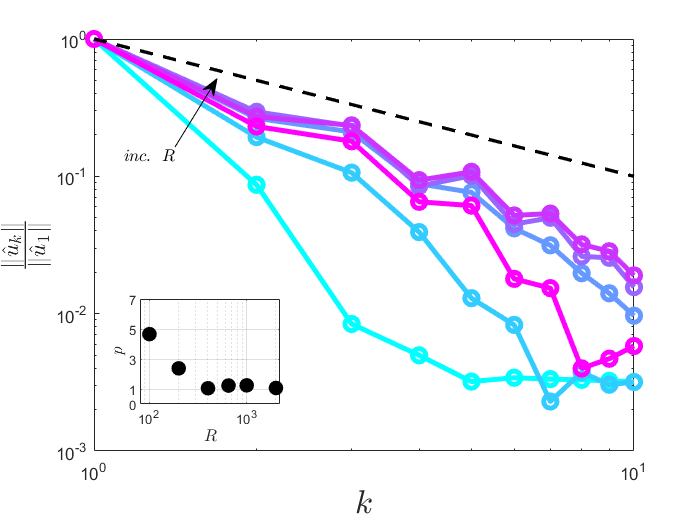}
    \caption{Norm of the Fourier modes computed from DNS at $R=100, ~\ 200, ~\ 400, ~\ 650, ~\ 1000, ~\ 2000$. Dashed black line is $\sim k^{-1}$. Inset shows exponent of best fit power law as in (\ref{decay power law}). Power law fit performed over the range $1<k<5$ for $R=100$ and $1<k<10$ for all $R>100$.}
    \label{fig:fourier_decay}
\end{figure}

Inspection of the spectral dynamics of the flow corroborate this finding. If we assume that the Fourier modes $\hat{\boldsymbol{u}}_k$ obey a power law
\begin{equation}\label{decay power law}
    \|\hat{\boldsymbol{u}}_k\| \sim k^{-p},
\end{equation}
then, from (\ref{f_cancel}), the forcing component $\hat{\boldsymbol{f}}^{\pm}_{k,cancel}$ must obey the power law
\begin{equation}
    \|\hat{\boldsymbol{f}}^{\pm}_{k,cancel}\| \sim k^{1-p}.
\end{equation}
Thus the flow will be in the WNL regime as long as $p\gg1$ and we expect the flow to have transitioned to the FNL regime if $p \lesssim 1$. In figure \ref{fig:fourier_decay} we plot the norm of the Fourier modes computed from the DNS data for a range of Reynolds numbers. For all cases the Fourier modes decay in Fourier space faster than $k^{-1}$ which is depicted by the dashed black line. However, the decay rate at $R=100$ is significantly faster than for the higher Reynolds number cases which seem to converge to a decay rate which is roughly independent of Reynolds number. The inset of figure \ref{fig:fourier_decay} shows the exponent of the best fit power law for all Reynolds numbers. For $R=100$ we fit the power law only to $k\leq5$ since for $5<k\leq10$ the norm of the Fourier components remains roughly constant. At $R=100$, in the WNL regime, the best fit exponent is approximately $5$ while the higher Reynolds numbers, which are in the FNL regime, all exhibit an exponent which seems to approach an asymptote close to $1$. These findings are in agreement with the analysis presented above which predicts that in the WNL regime the decay rate of the Fourier modes is much faster than $k^{-1}$ and the transition to the fully nonlinear regime is associated with the decay rate approaching $k^{-1}$.

\subsection{Model reduction}\label{sec:model reduction}
Here we address how the particular truncation values $N_{SVD}^k$ were chosen, and how the number of retained wavenumbers and resolvent modes at each wavenumber affects the accuracy of the model. At $R=100$ the flow is in the weakly nonlinear regime and thus the flow may be arbitrarily truncated in Fourier space with out appreciably impacting the accuracy of the retained harmonics. Additionally, in this case the optimal resolvent mode is a good approximation of the flow and thus retaining only a single harmonic with $N_{SVD}^1 = 2$ and  $N_{SVD}^2 = 5$ is sufficient to converge to a result whose 2D representation  (figures \ref{fig:2d result u} and \ref{fig:2d result vort}) is visually indistinguishable from the DNS. However, we retain more wavenumbers and resolvent modes than this in the results discussed in \S\ref{sec:nonlinear} in order to highlight the structure of the nonlinear forcing. At this Reynolds number the increase in computational cost to do so is trivial. 

\begin{figure}
    \centering
    \includegraphics[trim = 200 95 0 5,scale=0.25]{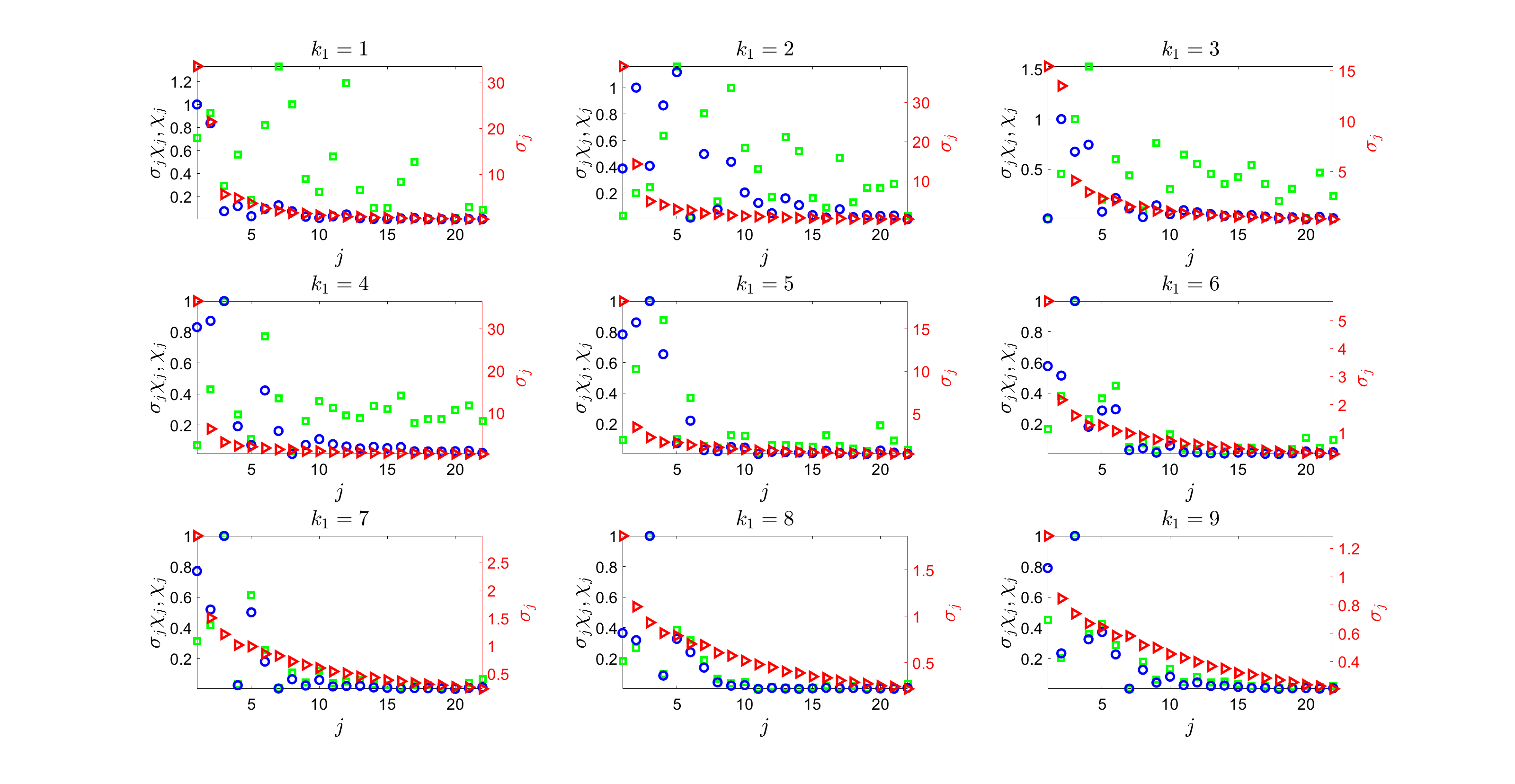}
    \caption{Expansion coefficients of the velocity $\sigma_{k,j}\chi_{k,j}$ (blue circles) and nonlinear forcing $\chi_{k,j}$(green squares), and singular values $\sigma_{k,j}$ (red triangles) for $R=400$.  Expansion coefficients are normalized by their maximum value at a given wavenumber and plotted against the left y-axis. Singular values are plotted against the right y-axis. Top row $k=1-3$, middle row $k=4-6$, bottom row: $k=7-9$. }
    \label{fig:sigma chi full}
\end{figure}

For the results in the fully nonlinear regime we focus the discussion here on $R=400$, with analogous arguments relevant to $R=200$. To establish a sufficiently converged baseline case from which to reduce the model complexity we increased $N_k$ and $N_{SVD}^k$ uniformly until the residual no longer decreased appreciably with added degrees of freedom. For $R=400$ this ``full" convergence was achieved with $N_k=9$ and $N_{SVD}^k=22$. In figure \ref{fig:sigma chi full} we plot the expansion coefficients $\sigma_{k,j}\chi_{k,j}$ in (\ref{uexp}) and the $\chi_{k,j}$ in (\ref{fexp}). The $\sigma_{k,j}\chi_{k,j}$ and $\chi_{k,j}$ represent the projection of the velocity and nonlinear forcing on to their respective resolvent basis $\boldsymbol{\psi}_{k,j}$ and $\boldsymbol{\phi}_{k,j}$ respectively. We also plot the singular values $\sigma_{k,j}$ on the right y-axis. Figure \ref{fig:sigma chi full} reveals that while the velocity is represented by only a few response modes, the nonlinear forcing has a significant projection onto many suboptimal modes.

This finding is in agreement with \citet{symon_energy_2021} and \citet{morra_colour_2021} who show that the nonlinear forcing has significant projection onto the sub-optimal resolvent forcing modes for a variety of flows even if the resolvent operator is low rank. The former considered both ECS as well as flow in a minimal channel while the latter focused entirely on turbulent channel flow. In fact, as also observed by \citet{morra_colour_2021}, the projection onto the first two forcing modes, $\chi_{k,1}$ and $\chi_{k,2}$, is much lower than the projection onto many of the suboptimal modes.

We interpret this significant projection of the nonlinear forcing onto a wide range of suboptimal modes as being necessitated by the roll off in singular values and the structured nature of the nonlinear forcing. If the nonlinear forcing was unstructured white noise, the solution would be given by the sum of resolvent response modes weighted by their singular values. However, for flows with structured nonlinear forcing, such as the TVF discussed here, the projection of the velocity field onto the response basis, $\boldsymbol{\psi}_{k,j}$ does not necessarily decrease monotonically with $j$ as evidenced by figure \ref{fig:sigma chi full}. Therefore, if we consider a certain suboptimal $\boldsymbol{\psi}_{k,j}$ which has a significant projection onto the velocity field, $\sigma_{k,j}\chi_{k,j}$ but is associated with a small singular value $\sigma_{k,j}$ the nonlinear forcing must by definition have a large projection onto the associated $\boldsymbol{\phi}_{k,j}$  since if $\sigma_{k,j}\chi_{k,j}$ is significant and $\sigma_{k,j}$ is small, then $\chi_{k,j}$ must be large. Thus the lack of roll off in the $\chi_{k,j}$ is a measure of how much the structure of the forcing differs from white noise.


From a practical point of view we see that for $2\leq k \leq 4$ the model solution has significant projection onto the majority of the retained singular response modes. The projections of the fundamental $(k=1)$ and the higher harmonics, $k>4$ generally have decayed to negligible levels for $j \gtrsim 10$. Neglecting these suboptimal modes for the higher harmonics does not affect the accuracy of the solution and the associated $30\%$ reduction in degrees of freedom results in a $90\%$ reduction in computational complexity since cost of computing the $6^{th}$ order tensors in (\ref{optprob1}) scale as $N^6$. Neglecting these negligible sub-optimal modes in the higher harmonics we arrive at the final truncation values cited in \S\ref{sec:math}, $N_{SVD}^k = 22 ~\ \forall k \leq 4,~\ N_{SVD}^k = 10~\ \forall k>4$. With this reduction in degrees of freedom the computational complexity has decreased to a point where the optimization may be carried out cheaply on a personal computer. It is these results using these values of $N_{SVD}^k$ that are presented in \S\ref{sec:results} and \S\ref{sec:nonlinear}. However, if only the large scales are desired or lower levels of convergence are acceptable, the solution is robust to significantly more truncation in both Fourier space and the SVD.   

\section{Efficient initial conditions for DNS}\label{sec:DNSic}
It is well known, if not entirely understood, that Taylor vortices persist well into the turbulent regime \citep{grossmann_highreynolds_2016}. While the nature of the Taylor vortices does evolve with increasing Reynolds number as discussed in this work and \citet{sacco_dynamics_2019}, the general structure does not deviate significantly from the form at $R=400$ shown in figures \ref{fig:2d result u} and \ref{fig:2d result vort}. Given the  significant model reduction achieved by our model, we now investigate whether the large scale Taylor vortex structure can be precomputed using our approach and then used to initialize a DNS at a higher Reynolds number to reduce the time to converge to a statistically stationary state. Similar ideas have been investigated by \citet{rosales_minimal_2006}, who initialized DNS and LES of isotropic decaying turbulence with both standard Gaussian and more realistic non-Gaussian vector fields. They found that the latter, which displayed some of the physical features associated with turbulence, led to shorter transition times before realistic decay rates were observed.

We performed two sets of DNS of Taylor Couette flow for a range of Reynolds numbers from $400$ to $2000$: the first using a random perturbation as an initial condition and one using the $R=400$ model solution as an initial condition. The simulations were run until the torque at the inner and outer cylinder agree to within $1\%$. The simulation was then continued for an additional 200 non-dimensional time units at which point the simulation was deemed to be converged. We define the percent reduction in time to convergence between the two cases
\begin{equation}
    P\equiv \frac{T_{0}-T_{m}}{T_{0}}\times100\%
\end{equation}
where $T_0$ and $T_m$ are the time required to reach convergence with the random and model initial conditions respectively. Table \ref{tab:time save} summarizes the savings for all the Reynolds numbers we considered. As expected the percentage of run time saved decreases as the Reynolds number increases because the Taylor vortices change slightly and, more crucially, because the flow becomes more three-dimensional and time dependent. However, it is remarkable that even at the highest Reynolds number, $R=2000$, which is five times the Reynolds number of the model used as an initial condition, the run time is reduced by $65\%$. Physically, this finding speaks to the robustness of the Taylor vortices, a phenomenon which has been observed by a host of authors \citep{grossmann_highreynolds_2016}. Practically, this suggests that high Reynolds number simulations of flows with large scale coherent structures may be made more cost effective by precomputing these large scale structures using a reduced order model such as the one presented here. 

\begin{table}
    \centering
    \begin{tabular}{c|c|c|c|c|c}
        $R$ & 400 & 650 & 1000 & 1500 & 2000  \\
        $P$ & $82\%$ & $80\%$ & $75\%$ & $72\%$ & $65\%$ 
    \end{tabular}
    \caption{Percentage reduction in convergence time using model TVF solution as initial condition compared to random perturbation as a function of Reynolds number. All cases use the $R=400$ model result as an initial condition.}
    \label{tab:time save}
\end{table}

\section{Conclusions}\label{sec:conclusion}
We have presented a fully nonlinear reduced order model of Taylor vortex flow for Reynolds numbers up to five times greater than the critical value. The resolvent formulation allows the governing equations for the fluctuations about a known mean velocity to be transformed into a set of polynomial equations. We approximate the solution to these equations by minimizing their associated residual in conjunction with a constraint which ensures the model generates Reynolds stresses compatible with the input mean velocity profile. We are able to generate model solutions which solve the NSE to a very good approximation and replicate the flow field computed through DNS at a tiny fraction of the computational cost. We believe this is the first explicit example of  ``closing the resolvent loop" published in the literature, although \citet{rosenberg_resolvent-based_2018} presented a similar analysis applied to ECS in a channel in his doctoral thesis which inspired this work. 
We analyzed the nonlinear interactions driving the flow for a range of Reynolds numbers and identified the transition from a weakly nonlinear regime close to the bifurcation from the laminar state where the structure of the flow is accurately modeled by the linear dynamics and the forcing cascade is purely from large scales to small scales. At higher Reynolds numbers we define a fully nonlinear regime where an inverse forcing cascade from small to large scales emerges to counter the cascade from large to small scales. The dominant nonlinear interactions at a given wavenumber $k$ involve the pair of triadic interactions  $k = (k\pm1) \mp 1$. For the highest Reynolds number case the pair of triads $k = (k\pm2) \mp 2$ also emerges as a dominant forcing mechanism. The velocity contributions from the pairs of triads involving the fundamental have opposite sign and almost equal amplitudes which are much larger than the full Fourier mode. Their sum comprises significant destructive interference with the small differences in shape giving rise to the shape of the full Fourier mode. We demonstrated that this destructive interference is a direct consequence of the structure of the nonlinear term of the NSE formulated in Fourier space. Furthermore, this bidirectional forcing cascade implies that in order to accurately model a flow up to a certain order in Fourier space significantly more harmonics than desired must be retained in order to capture this inverse forcing cascade. We postulated that this shift from linear/weakly nonlinear to fully nonlinear dynamics is related to a similar transition in the physics of Taylor vortices observed by \citet{sacco_dynamics_2019}.

Finally, we used our model solution as an initial condition to DNS of TCF at higher Reynolds numbers and were able to significantly reduce the time to convergence compared to initializing the simulation with a random perturbation. These results suggest that simulation of flows with large scale coherent structures may improved by precomputing these coherent structures using a reduced order model.

\section{Acknowledgments}
We thank Kevin Rosenberg for many inspiring and helpful discussions and gratefully acknowledge the support of the U.S. Office of Naval Research under grants ONR N00014-17-1-2307 and N00014-17-1-3022.
\appendix
\section{}\label{appA}
The Navier-Stokes operator in cylindrical coordinates linearized about a one dimensional azimuthal mean flow $U(r)$  and Fourier transformed in $\theta, z,$ and $t$ is given by
\begin{equation}\label{L}
    \setlength{\arraycolsep}{8pt}
\renewcommand{\arraystretch}{2.0}
\mathcal{L}_{\mathbf{k}} = \left[
\begin{array}{cccc}
    \frac{i n U}{r} + \frac{1}{R}\left(\frac{1}{r^2}-\nabla^2\right)  &  \frac{1}{R}\left(\frac{2in}{r^2}\right) - \frac{2U}{r}  &  0  &  \frac{\partial}{\partial r}  \\
  
    \left(\frac{\partial U}{\partial r} + \frac{U}{r}\right) - \frac{1}{R}\left(\frac{2in}{r^2}\right)  &  \frac{i n U}{r} + \frac{1}{R}\left(\frac{1}{r^2}-\nabla^2 \right)   &  0  &  \frac{in}{r}   \\
    
    0  &  0  &  \frac{i n U}{r} - \frac{1}{R}\nabla^2  &  i k    \\
 
    \frac{1}{r} + \frac{\partial}{\partial r}  &  \frac{i n}{r}  &  i k  &  0

\end{array}  \right],
\end{equation}
where the Laplacian operator is defined as
\begin{equation}
    \nabla^2 =   \frac{\partial^2}{\partial r^2} + \frac{1}{r}\frac{\partial}{\partial r} -\left(k^2 + \frac{n^2}{r^2}  \right).
\end{equation}
The weight matrix, $M$, is defined as 
\begin{equation}
M \equiv
\left[
    \begin{array}{cccc}
        1 & 0 & 0 & 0  \\
         0 & 1 & 0 & 0  \\
         0 & 0 & 1 & 0  \\
         0 & 0 & 0 & 0  \\ 
    \end{array}
     \right].
\end{equation}

For the streamwise constant modes considered here the continuity equation may be used to eliminate pressure and the axial velocity to write the system in terms of the radial and azimuthal velocity, $\hat{\boldsymbol{q}}=[\hat{u}_r,\hat{u}_{\theta}]$ such that 
\begin{equation}
    \Tilde{\mathcal{L}_k} \hat{\boldsymbol{q}} = \Tilde{M}\hat{\boldsymbol{f}}
\end{equation}
where $\hat{\boldsymbol{f}} = [\hat{f}_r,\hat{f}_{\theta},\hat{f}_z]$ and the linear operators take the form
\begin{equation}\label{Lk1}
    \Tilde{\mathcal{L}}_k = 
    \begin{bmatrix}
         -\frac{1}{R}\nabla^4 & \frac{2k^2\overline{U}}{r}\\
         \frac{1}{r}\frac{\partial}{\partial r}\left(r\overline{U}\right) & -\frac{1}{R}\nabla^2 
    \end{bmatrix},
\end{equation}

\begin{equation}
\Tilde{M} \equiv
\left[
    \begin{array}{ccc}
         -k_z^2& 0 & -ik_z \frac{\partial}{\partial r}  \\
         0 & 1 & 0   \\

    \end{array}
     \right].
\end{equation}

\newpage

\bibliography{references2.bib}

\begin{thebibliography}{}

\bibitem[Andereck et~al., 1986]{andereck_flow_1986}
Andereck, C.~D., Liu, S.~S., and Swinney, H.~L. (1986).
\newblock Flow regimes in a circular {Couette} system with independently
  rotating cylinders.
\newblock {\em Journal of Fluid Mechanics}, 164:155--183.

\bibitem[Beaume et~al., 2015]{beaume_reduced_2015}
Beaume, C., Chini, G.~P., Julien, K., and Knobloch, E. (2015).
\newblock Reduced description of exact coherent states in parallel shear flows.
\newblock {\em Physical Review E}, 91(4):043010.

\bibitem[Coles, 1965]{coles_transition_1965}
Coles, D. (1965).
\newblock Transition in circular {Couette} flow.
\newblock {\em Journal of Fluid Mechanics}, 21(3):385--425.

\bibitem[Dessup et~al., 2018]{dessup_self-sustaining_2018}
Dessup, T., Tuckerman, L.~S., Wesfreid, J.~E., Barkley, D., and Willis, A.~P.
  (2018).
\newblock Self-sustaining process in {Taylor}-{Couette} flow.
\newblock {\em Physical Review Fluids}, 3:123902.

\bibitem[Gallaire et~al., 2016]{gallaire_pushing_2016}
Gallaire, F., Boujo, E., Mantic-Lugo, V., Arratia, C., Thiria, B., and Meliga,
  P. (2016).
\newblock Pushing amplitude equations far from threshold: application to the
  supercritical {Hopf} bifurcation in the cylinder wake.
\newblock {\em Fluid Dynamics Research}, 48(6):061401.

\bibitem[Gebhardt and Grossmann, 1993]{gebhardt_taylor-couette_1993}
Gebhardt, T. and Grossmann, S. (1993).
\newblock The {Taylor}-{Couette} eigenvalue problem with independently rotating
  cylinders.
\newblock {\em Zeitschrift für Physik B Condensed Matter}, 90(4):475--490.

\bibitem[Grossmann et~al., 2016]{grossmann_highreynolds_2016}
Grossmann, S., Lohse, D., and Sun, C. (2016).
\newblock High–{Reynolds} {Number} {Taylor}-{Couette} {Turbulence}.
\newblock {\em Annual Review of Fluid Mechanics}, 48:53--80.

\bibitem[Hamilton et~al., 1995]{hamilton_regeneration_1995}
Hamilton, J.~M., Kim, J., and Waleffe, F. (1995).
\newblock Regeneration mechanisms of near-wall turbulence structures.
\newblock {\em Journal of Fluid Mechanics}, 287:317--348.

\bibitem[Huisman et~al., 2014]{huisman_multiple_2014}
Huisman, S.~G., van~der Veen, R. C.~A., Sun, C., and Lohse, D. (2014).
\newblock Multiple states in highly turbulent {Taylor}-{Couette} flow.
\newblock {\em Nature Communications}, 5:3820.

\bibitem[Illingworth, 2020]{illingworth_streamwise-constant_2020}
Illingworth, S.~J. (2020).
\newblock Streamwise-constant large-scale structures in {Couette} and
  {Poiseuille} flows.
\newblock {\em Journal of Fluid Mechanics}, 889.

\bibitem[Jones, 1981]{jones_nonlinear_1981}
Jones, C.~A. (1981).
\newblock Nonlinear {Taylor} vortices and their stability.
\newblock {\em Journal of Fluid Mechanics}, 102:249--261.

\bibitem[Mantič-Lugo et~al., 2014]{mantic-lugo_self-consistent_2014}
Mantič-Lugo, V., Arratia, C., and Gallaire, F. (2014).
\newblock Self-{Consistent} {Mean} {Flow} {Description} of the {Nonlinear}
  {Saturation} of the {Vortex} {Shedding} in the {Cylinder} {Wake}.
\newblock {\em Physical Review Letters}, 113(8):084501.

\bibitem[Mantič-Lugo et~al., 2015]{mantic-lugo_self-consistent_2015}
Mantič-Lugo, V., Arratia, C., and Gallaire, F. (2015).
\newblock A self-consistent model for the saturation dynamics of the vortex
  shedding around the mean flow in the unstable cylinder wake.
\newblock {\em Physics of Fluids}, 27(7):074103.

\bibitem[Marcus, 1984]{marcus_simulation_1984}
Marcus, P.~S. (1984).
\newblock Simulation of {Taylor}-{Couette} flow. {Part} 2. {Numerical} results
  for wavy-vortex flow with one travelling wave.
\newblock {\em Journal of Fluid Mechanics}, 146:65--113.

\bibitem[Maretzke et~al., 2014]{maretzke_transient_2014}
Maretzke, S., Hof, B., and Avila, M. (2014).
\newblock Transient growth in linearly stable {Taylor}–{Couette} flows.
\newblock {\em Journal of Fluid Mechanics}, 742:254--290.

\bibitem[McKeon and Sharma, 2010]{mckeon_critical-layer_2010}
McKeon, B.~J. and Sharma, A.~S. (2010).
\newblock A critical-layer framework for turbulent pipe flow.
\newblock {\em Journal of Fluid Mechanics}, 658:336--382.

\bibitem[McMullen et~al., 2020]{mcmullen_interaction_2020}
McMullen, R.~M., Rosenberg, K., and McKeon, B.~J. (2020).
\newblock Interaction of forced {Orr}-{Sommerfeld} and {Squire} modes in a
  low-order representation of turbulent channel flow.
\newblock {\em Physical Review Fluids}, 5(8):084607.

\bibitem[Moarref et~al., 2014]{moarref_low-order_2014}
Moarref, R., Jovanović, M.~R., Tropp, J.~A., Sharma, A.~S., and McKeon, B.~J.
  (2014).
\newblock A low-order decomposition of turbulent channel flow via resolvent
  analysis and convex optimization.
\newblock {\em Physics of Fluids}, 26(5):051701.

\bibitem[Morra et~al., 2021]{morra_colour_2021}
Morra, P., Nogueira, P. A.~S., C., A. V.~G., and Henningson, D.~S. (2021).
\newblock The colour of forcing statistics in resolvent analyses of turbulent
  channel flows.
\newblock {\em Journal of Fluid Mechanics}, 907:A24.

\bibitem[Nagata, 1990]{nagata_three-dimensional_1990}
Nagata, M. (1990).
\newblock Three-dimensional finite-amplitude solutions in plane {Couette} flow:
  bifurcation from infinity.
\newblock {\em Journal of Fluid Mechanics}, 217:519--527.

\bibitem[Nogueira et~al., 2021]{nogueira_forcing_2021}
Nogueira, P. A.~S., Morra, P., Martini, E., Cavalieri, A. V.~G., and
  Henningson, D.~S. (2021).
\newblock Forcing statistics in resolvent analysis: application in minimal
  turbulent {Couette} flow.
\newblock {\em Journal of Fluid Mechanics}, 908.

\bibitem[Ostilla et~al., 2013]{ostilla_optimal_2013}
Ostilla, R., Stevens, R. J. A.~M., Grossmann, S., Verzicco, R., and Lohse, D.
  (2013).
\newblock Optimal {Taylor}–{Couette} flow: direct numerical simulations.
\newblock {\em Journal of Fluid Mechanics}, 719:14--46.

\bibitem[Ostilla-Mónico et~al., 2014]{ostilla-monico_exploring_2014}
Ostilla-Mónico, R., van~der Poel, E.~P., Verzicco, R., Grossmann, S., and
  Lohse, D. (2014).
\newblock Exploring the phase diagram of fully turbulent {Taylor}–{Couette}
  flow.
\newblock {\em Journal of Fluid Mechanics}, 761:1--26.

\bibitem[Rand, 1982]{rand_dynamics_1982}
Rand, D. (1982).
\newblock Dynamics and symmetry. {Predictions} for modulated waves in rotating
  fluids.
\newblock {\em Archive for Rational Mechanics and Analysis}, 79(1):1--37.

\bibitem[Rigas et~al., 2021]{rigas_nonlinear_2021}
Rigas, G., Sipp, D., and Colonius, T. (2021).
\newblock Nonlinear input/output analysis: application to boundary layer
  transition.
\newblock {\em Journal of Fluid Mechanics}, 911.

\bibitem[Rosales and Meneveau, 2006]{rosales_minimal_2006}
Rosales, C. and Meneveau, C. (2006).
\newblock A minimal multiscale {Lagrangian} map approach to synthesize
  non-{Gaussian} turbulent vector fields.
\newblock {\em Physics of Fluids}, 18(7):075104.

\bibitem[Rosenberg, 2018]{rosenberg_resolvent-based_2018}
Rosenberg, K. (2018).
\newblock {\em Resolvent-based modeling of ﬂows in a channel}.
\newblock PhD thesis, California Institute of Technology.

\bibitem[Rosenberg and McKeon, 2019a]{rosenberg_computing_2019}
Rosenberg, K. and McKeon, B.~J. (2019a).
\newblock Computing exact coherent states in channels starting from the laminar
  profile: {A} resolvent-based approach.
\newblock {\em Physical Review E}, 100(2):021101.

\bibitem[Rosenberg and McKeon, 2019b]{rosenberg_efficient_2019}
Rosenberg, K. and McKeon, B.~J. (2019b).
\newblock Efficient representation of exact coherent states of the
  {Navier}–{Stokes} equations using resolvent analysis.
\newblock {\em Fluid Dynamics Research}, 51(1):011401.

\bibitem[Sacco et~al., 2019]{sacco_dynamics_2019}
Sacco, F., Verzicco, R., and Ostilla-Mónico, R. (2019).
\newblock Dynamics and evolution of turbulent {Taylor} rolls.
\newblock {\em Journal of Fluid Mechanics}, 870:970--987.

\bibitem[Stuart, 1960]{stuart_non-linear_1960}
Stuart, J.~T. (1960).
\newblock On the non-linear mechanics of wave disturbances in stable and
  unstable parallel flows {Part} 1. {The} basic behaviour in plane {Poiseuille}
  flow.
\newblock {\em Journal of Fluid Mechanics}, 9(3):353--370.

\bibitem[Symon et~al., 2021]{symon_energy_2021}
Symon, S., Illingworth, S.~J., and Marusic, I. (2021).
\newblock Energy transfer in turbulent channel flows and implications for
  resolvent modelling.
\newblock {\em Journal of Fluid Mechanics}, 911.

\bibitem[Taylor, 1923]{taylor_viii_1923}
Taylor, G.~I. (1923).
\newblock {VIII}. {Stability} of a viscous liquid contained between two
  rotating cylinders.
\newblock {\em Philosophical Transactions of the Royal Society of London.
  Series A, Containing Papers of a Mathematical or Physical Character},
  223(605-615):289--343.

\bibitem[van~der Poel et~al., 2015]{van_der_poel_pencil_2015}
van~der Poel, E.~P., Ostilla-Mónico, R., Donners, J., and Verzicco, R. (2015).
\newblock A pencil distributed finite difference code for strongly turbulent
  wall-bounded flows.
\newblock {\em Computers \& Fluids}, 116:10--16.

\bibitem[van Gils et~al., 2011]{van_gils_twente_2011}
van Gils, D. P.~M., Bruggert, G., Lathrop, D.~P., Sun, C., and Lohse, D.
  (2011).
\newblock The {Twente} turbulent {Taylor}-{Couette} ({T3C}) facility: strongly
  turbulent (multiphase) flow between two independently rotating cylinders.
\newblock {\em The Review of Scientific Instruments}, 82(2):025105.

\bibitem[van Gils et~al., 2012]{van_gils_optimal_2012}
van Gils, D. P.~M., Huisman, S.~G., Grossmann, S., Sun, C., and Lohse, D.
  (2012).
\newblock Optimal {Taylor}–{Couette} turbulence.
\newblock {\em Journal of Fluid Mechanics}, 706:118--149.

\bibitem[Verzicco and Orlandi, 1996]{verzicco_finite-difference_1996}
Verzicco, R. and Orlandi, P. (1996).
\newblock A {Finite}-{Difference} {Scheme} for {Three}-{Dimensional}
  {Incompressible} {Flows} in {Cylindrical} {Coordinates}.
\newblock {\em Journal of Computational Physics}, 123(2):402--414.

\bibitem[Waleffe, 1997]{waleffe_self-sustaining_1997}
Waleffe, F. (1997).
\newblock On a self-sustaining process in shear flows.
\newblock {\em Physics of Fluids}, 9(4):883--900.

\bibitem[Yahata, 1977]{yahata_slowly-varying_1977}
Yahata, H. (1977).
\newblock Slowly-{Varying} {Amplitude} of the {Taylor} {Vortices} near the
  {Instability} {Point}. {II}: {Mode}-{Coupling}-{Theoretical} {Approach}.
\newblock {\em Progress of Theoretical Physics}, 57(5):1490--1496.

\bibitem[Zhu et~al., 2018]{zhu_afid-gpu_2018}
Zhu, X., Phillips, E., Spandan, V., Donners, J., Ruetsch, G., Romero, J.,
  Ostilla-Mónico, R., Yang, Y., Lohse, D., Verzicco, R., Fatica, M., and
  Stevens, R. J. A.~M. (2018).
\newblock {AFiD}-{GPU}: {A} versatile {Navier}–{Stokes} solver for
  wall-bounded turbulent flows on {GPU} clusters.
\newblock {\em Computer Physics Communications}, 229:199--210.

\end{thebibliography}

\end{document}